# Hidden information and regularities of information dynamics 1R

*Vladimir S. Lerner*
13603 Marina Pointe Drive, Marina Del Rey, CA 90292, USA lernervs@gmail.com

*Abstract*

The introduced entropy functional's (EF) information measure of a random process integrates all information contribution along the process trajectories, evaluating both the states' and between states' bound information connections.

By defining the integral information measure on the trajectories of the multi-dimensional Markov stochastic dynamics (as a simple mathematical model of a random nonstationary interactive process), we obtain the formal integral evaluation of the process that includes its multiple long-term correlations. This measure reveals information that is hidden by traditional information measures, which commonly use Shannon's entropy function for each selected stationary states of the process.

Such hidden information could disclose a missing connections during, for example, creation of a human thought, speech, discussion, communication, and so on, being important for evaluation the process' meaningful information, related to these connections in acceptance, cognition, and perception, which enables producing a logic of the information.

The presentation consists of three Parts. In Part1R (revised) we analyze a *mechanism* of arising information regularities from a stochastic process, measured by EF, independently of the process' specific source and origin, applying them to a wide class of real material and non material processes, including an intellectual process and a world of virtual reality.

Uncovering the process' *regularities* leads us to an *information law*, based on *extracting a maximal information from its minimum*, which could create these regularities. To formally approach such a law, we solve a variation problem (VP) of converting a stochastic process, measured by EF with its hidden information, to a dynamic process, measured by an information path functional (IPF) on the process' trajectories. The IPF variation equations determine the information dynamic model, approximating the stochastic process with a maximal functional probability on trajectories and the equations for a joint solution of the identification and optimal control problems, combined with a state's consolidation. The VP information invariants allow developing a simple computer based procedure for dynamic modeling and prediction of the random diffusion process with following optimal encoding of the identified dynamic model's operator and control. In Part 2, we study the cooperative processes, arising at the consolidation, as a result of the VP-EF-IPF approach, which is able to produce a multiple cooperative structures, concurrently assembling in a hierarchical information network (IN) and generating the IN's digital genetic code. We analyze information geometry of the cooperative structures, evaluate a curvature of their geometrical forms and their cooperative information complexities.

The dynamic model, which extracts a hidden information of an observed process, does both conversion, cooperation, and then generation of its genetic code, working as an information *operating* system which *creates an information observer*.

In Part 3 we study the evolutionary information processes and regularities of *evolution dynamics*, evaluated by entropy functional (EF) of a random field and informational path functional (IPF) of a dynamic space-time process.

The VP minimax variation principle, applied to the evolution of both random microprocesses and dynamic macroprocesses, leads to the following evolutionary regularities: creation of the order from stochastics through the evolutionary macrodynamics, described by a gradient of dynamic potential, evolutionary speed and the evolutionary conditions of a fitness and diversity; the evolutionary hierarchy with growing information values and potential adaptation; the adaptive self-controls and a self-organization with a mechanism of copying to a genetic code.

This law and the regularities determine unified functional informational mechanisms of evolution dynamics.

*Introduction*

Conventional information science considers an information process, but traditionally uses the probability measure for random states and Shannon's entropy as the uncertainty function of these states.

We present a new approach, based on an *integral information measure* of the random process, which extends Shannon's entropy measure. The integral information measure, applied to the whole process, reveals *hidden dynamic* connections between the process' states and allows to describe the process' *dynamics* and disclose its regularities.

Such a measure could evaluate the process' *meaningful* information related to the acceptance, cognition, and perception of the information. For example, a pattern of hidden information could reveal a missing essential connections in a human perception. Spoken language with its many phonemes contains the extra variables, which are used in classical non-redundant encoding, allocating some redundant bits for error corrections. Redundancy, built in any natural language, as its background, allows people better understand conversation than the related written text. The letters' words' phrase's connections through their correlations could be lost at translation, writing, encoding (which usually does not include the short and /or extended correlations). These multiple connections covered via the correlations, being integrated throughout understanding of their meaning, create human thought, ideas.

That is why information, measured by classical Shannon's entropy, aimed on optimal encoding information process' states, contains less information than a natural conversation, as a whole process, before being encoded.

The considered functional information measure that integrates all process, including all inner connections of the process states, holds more information than the sum of Shannon's measures for the states, and, hence, can be constructive for revealing both meaning and idea of via this information.

This approach aimed on a formal measuring of both a human understanding and semantics of the information.

We define an information process as a logical structure of mutually connected information symbols, measured by the information integral, which takes into account its inter-symbol's statistical dependencies.

We analyze the information process' regularities *independently* of the process' specific *source and origin*, applying them to a wide class of real material and non material processes, including an intellectual process and a world of virtual reality. Such a formal approach allows us to reveal some *general* features and regularities of information process and information structure generated by the process' interactive dynamics. To uncover the process' regularities we study an *information law*, based on *extracting a maximal information from its minimum*. The law minimum generates optimal information dynamics of the collected information, which could create these *regularities*. The approach could be useful for revealing a specifc form of the *information law* for the observed random process and its important phenomena. The law allows us to predict a multiple interactive dynamics of the process' information flows, including their collectivization in complex information systems structures, information networks with a specific time–space information structures and optimal information *code*, and also regularities of the process' evolutionary dynamics. Such a law, applied to an intellect, describes regularities of information flows at acquisition and accumulation of intellectual information. The law, applied to a *multiple* intellect, allows describing their united intellect, while their dissimilarity depends on each intellect's information, being previously collected, acquired, and accumulated, which determines the intellect's information code.

Multiple intellect's regularities could change each of the individual's code intellect, creating their united code.

Knowing information law's regularities leads to *predictable evolution dynamics of information processes.*

**Part I. Information Macrodynamic Approach. The main formalism**

*1. Notion of information applied to a process*

On intuitive level, information is associated with diverse forms of *changes* (transformations) in material and/or non-material objects of observation, expressed *universally and* unconnectedly to the changes' cause and origin.

Such generalized changes are brought potentially by *random* events and processes as a set, described via the related probability in a *probability space*, which are studied in the theory of probability, founded as a *logical* science (Kolmogorov 1968). This randomness with their probabilities we consider as a *source* of information, which implies that some of them, but not all of such randomness produces information.

The source specifics depend on its particular random events in the *probability space.*

*The questions* are: How to describe a change formally and evaluate information generated by the change?

We consider generally a random process (as a continuous or discrete function $x(\omega, s)$ of random variable $\omega$ and time $s$), described by the *changes* of its elementary probabilities from one distribution (a *priory*) $P_{s,x}^a(d\omega)$ to another distribution (*a posteriori*) $P_{s,x}^p(d\omega)$ in the form of their transformation

$$p(\omega) = \frac{P_{s,x}^a(d\omega)}{P_{s,x}^p(d\omega)}. \tag{1.1}$$

Such a probabilistic description *generalizes* different forms of specific functional relations, represented by a sequence of different transformations, which might be extracted from the source, using the probabilities ratios (1.1).

It is convenient to measure the probability ratio in the form of natural logarithms: $\ln[p(\omega)]$, where the logarithm of each probability $\leq 1$ would be negative, or zero.

That is why the considered function holds the form of a logarithmical distance

$$-\ln p(\omega) = -\ln P_{s,x}^a(d\omega) - (-\ln P_{s,x}^p(d\omega)) = s_a - s_p = s_{ap}, \tag{1.1a}$$

represented by a difference of a *priory* $s_a > 0$ and a *posteriori* $s_p > 0$ *entropies*, each of them measures uncertainty for given $x(\omega, s)$ in the related distributions, while the difference measures uncertainty at the transformation of probabilities for the source events, which satisfies the entropy's additivity.

A *change* brings a certainty or *information* if its uncertainty $s_{ap}$ is removed by some equivalent entity call information $i_{ap} : s_{ap} - i_{ap} = 0$. Thus, information is delivered if $s_{ap} = i_{ap} > 0$, which requires $s_p < s_a$ and a positive logarithmic measure with $0 < p(\omega) < 1$. Condition of zero information: $i_{ap} = 0$ corresponds to a redundant change, transforming a priory probability to the equal a posteriori probability, or this transformation is an identical–informational *undistinguished*, redundant. In the same manner, each of the above entropies can be turned into related information.

The removal of uncertainty $s_a$ by $i_a : s_a - i_a = 0$ brings an equivalent certainty or *information* $i_a$ about entropy $s_a$.

In addition, a *posteriori uncertainty* decreases (or might remove) a priory uncertainty. If $s_p = s_a$, then $s_p$ brings information $i_p = i_a$ regarding $s_a$ by removing this uncertainty, but such a change does not deliver information.

At $s_p < s_a$, the $s_p$ still brings some information about $s_a$ but holds a non zero uncertainty $s_a - s_p = s_{ap} \geq 0$, which can be removed by the considered equivalent certainty-information $i_{ap}$. Uncertainty $s_a$ could also be a result of some a priory transformation, as well as uncertainty $s_p$ is resulted from a posteriori transformation.

Because each above probability and entropy is random, to find an average tendency of functional relations (1.1,1.1a) it is applied a mathematical expectation:

$$E_{s,x}\{-\ln[p(\omega)]\} = E_{s,x}[s_{ap}] = S_{ap} = I_{ap} \neq 0, \qquad (1.2)$$

which we call the mean *information of a random source*, being averaged by the source *events* (a probability state), or by the source *processes*, depending on what it is considered: a process, or an event, since both (1.2) and (1.1) include also Shannon's formula for information of a states (events) (Shannon, Weaver 1949).

Indeed. The mathematical expectation (1.2), applied to a process' differential probabilities $P_{\tilde{x}}^a$ and $P_{\tilde{x}}^p$ in (1.1a), considered along the process trajectories, acquires the form of an *entropy functional*

$$I_{ap} = -E_{s,x}[\ln(P_{s,x}^a / P_{s,x}^p)] = -\int_X \ln(P_{s,x}^a / P_{s,x}^p) P_{s,x}^a (d\tilde{x}_t), \qquad (1.2a)$$

where a transformed a posteriori process $\tilde{x}_t^p$, satisfying to the information transformation, we call *information process*, whose logical structure depends on the Kolmogorov probabilities. Information generally evaluates various multiple *relationships* represented via transformations (1.1-1.2) and generalizes them, being *independent* on diverse physical entities that carry this information. Some *logical* transformations in symbolic dynamic systems theory preserve an entropy as a metric invariant, allowing to classify these transformations. Therefore, *information* constitutes a *universal nature and a general measure of a transformation*, which conveys the *changes* decreasing uncertainty.

Having a *source* of information (data, symbols, relations, links, etc.) does not mean we get *information* and might evaluate its quantity and or a quality. A source only provides changes, whose information is measured by (1.2), (1.2a).

To obtain information, a subset of probability space, selected by formula (1.2) from the source set of probability space, should *not be empty*, which is measured by $I_{ap} \neq 0$.

*Definition 1*. Information, selected by formula (1.1a) from the source set of probability space, is *defined* by *not an empty subset of probability space*, which chooses only *a non-repeating (novel) subset* from the source.

*Definition 2*. Numerical value (in Nat, or Bit), measured by formula (1.2), determines the *quantity* of information selected from the source.

While, the notion of information formally separates the *distinguished from the undistinguished* (repeating) subsets (events, processes), formula (1.2) *evaluates numerically this separation*. The *information process*, satisfying such information measure (in form (1.2a)) selects no repeating processes from a source.

From these definitions it follows that both information, as a *subset* (a string) in probability space, and its measure are *mathematical* entities, *resulting* generally from logical operations, while both *delivering* a source and the *selection* from the source *do not define* notion of information. In our approach, these actions (operations) are performed by an *information observer*, which, first, selects the uncertain process by its information measure (1.2a) and then, converts it to a certain process, considered below. Some of operations, as well as transmission and acquisition of information, require spending energy, that leads to binding information with energy, and to the consideration of related physical, thermodynamic substances, associated with this conversion. A classical physical measurement of this information value (or a "barrier" of the above separation) requires an energy cost to decrease the related physical entropy at least by constant $k_B(\Delta T)$ (depending on a difference of related temperatures $\Delta T$), which transforms the information measure to the physical one, at measuring, or erasing and memorizing this information .

A quantum measurement involves the collapse of a wave function needed to transform the information *barrier*, separating the distinguished from the undistinguished subsets, to a quantum physical level, which corresponds to an interaction, or an observation. In both cases, such operations bind the required equivalent energy with information.

The mathematical definition of information (1.2) can be used to define its mathematical equivalent of energy:

$$E_{ap} = I_{ap} k_B . \qquad (1.2b)$$

which here is *asymmetrical* as its *integral* information measure (1.2).

Since probabilities can be defined for both non material and material substances, transformation (1.1) connects them in the form of a relative *information* (1.2), or (1.2a), which serves as both a *bridge* between diverse entities and a *common measure* of their connections.

However, there are no reversible equivalent relations between energy-information: the same quantity of information could be produced with different cost of energy, and vice versa, the same energy can produce different quantity of information.

The information subset, following from the definitions, is preserved and standardized by encoding this information in different form of information language, which assigns a specific code-word from some chosen language's alphabet to each sequence of the subset symbols. An *optimal* encoding allows transforming the encoded information with preservation of its information quantity. The encoding of different observed processes by a common (an optimal) code brings a unique opportunity of revealing the process' connections through its information measure and by encoding the process' specific feature in the related specific code. For example, DNA code encodes amino acids, which establishes the connection between biological nature of the encoded process and its information measure in bites.

The length of the shortest representation of the information string by a *code* or a *program* is the subject of *algorithmic information theory* (AIT). "AIT is the result of putting Shannon's information theory and Turing's computability theory to measure the complexity of an object by the size in bits of the smallest program for computing it" (Chaitin 1987). Encoding a process by the integral information measure (Lerner 2009), (Sec.2) provides a shortest finite program (in bits), connecting the considered approach to AIT and to Kolmogorov's complexity (See also Part 2).

## 2. Information functional measure of a Markov diffusion process

The process' simple example is a Brownian interactive movement, Levy walks, others, which models many physical, chemical and biological phenomena, and has important applications in economics and finances.

A diffusion process is defined as a continuous Markov process, satisfying, generally, a stochastic $n$-dimensional controllable differential equations in Ito's form:

$$d\tilde{x}_t = a(t, \tilde{x}_t, u_t)dt + \sigma(t, \tilde{x}_t)d\xi_t, \tilde{x}_s = \eta, t \in [s,T] = \Delta, s \in [0,T] \subset R_+^1, \qquad (2.1)$$

with the standard limitations (Dynkin 1979) on the functions of a controlled drift $a(t, \tilde{x}_t, u_t)$, diffusion $\sigma(t, \tilde{x}_t)$, and Wiener process $\xi_t = \xi(t, \omega)$, which are defined on a probability space of the elementary random events $\omega \in \Omega$ with the variables located in $R^n$; $\tilde{x}_t = \tilde{x}(t)$ are solutions of (2.1)) under applied control $u_t$. (The drift and diffusion functions are defined through the process' probability and the solutions of (2.1)(Prochorov, Rozanov 1973).

Suppose that the control function $u_t$ provides the transformation of a priory $P_{s,x}^a(d\omega)$ to a posteriori probability $P_{s,x}^p(d\omega)$, where a priory process $\tilde{x}_t^a$ is a solution of (2.1) prior to applying this control, at $a(t, \tilde{x}_t, u_t) = 0$, and a posteriori process $\tilde{x}_t^p$ is a solution of (2.1) after such a control provides this transformation, at $a(t, \tilde{x}_t, u_t) \neq 0$.

Such a priori process $\tilde{x}_t^a = \int_s^t \sigma(v, \zeta_v)d\zeta_v$ models an uncontrollable noise with $E[\tilde{x}_t^a] = O$. (2.1a)

The process' $\tilde{x}_t^p$ transformed probability is defined through its transition probability

$$P^p(s,x,t,B) = \int_{x_t \in B} (p(\omega))^{-1} P_{s,x}^a(d\omega) \qquad (2.2)$$

with a probability density measure (1.1) of this transformation in the form

$$p(\omega) = \frac{P_{s,x}^a(d\omega)}{P_{s,x}^p(d\omega)} = \exp\{-\varphi_s^t(\omega)\}, \qquad (2.3)$$

which for the above solutions of (2.1) is determined through the additive functional of the diffusion process (Dynkin1960):

$$\varphi_s^T = 1/2 \int_s^T a^u(t,\tilde{x}_t)^T (2b(t,\tilde{x}_t))^{-1} a^u(t,\tilde{x}_t) dt - \int_s^T (\sigma(t,\tilde{x}_t))^{-1} a^u(t,\tilde{x}_t) d\xi(t). \quad (2.4)$$

Using the definition of quantity information $I_{ap}$, obtained at this transformation, by entropy measure (1.2) for (2.3):

$$E_{s,x}\{-\ln[p(\omega)]\} = S_{ap} = I_{ap}, \; S_{ap} = E_{s,x}\{\varphi_s^t(\omega)\} \qquad (2.5)$$

and after substituting math expectation of (2.4) in (2.5) (at $E[\tilde{x}_t^a] = O$) we obtain entropy integral functional for a transformed process $\tilde{x}_t$:

$$S_{ap}[\tilde{x}_t]|_s^T = 1/2 E_{s,x}\{\int_s^T a^u(t,\tilde{x}_t)^T (2b(t,\tilde{x}_t))^{-1} a^u(t,\tilde{x}_t) dt\} = \int_{\tilde{x}(t) \in B} -\ln[p(\omega)] P_{s,x}^a(d\omega) = -E_{s,x}[\ln p(\omega)], \quad (2.6)$$

where $a^u(t,\tilde{x}_t) = a(t,\tilde{x}_t,u_t)$ is a drift function, depending on a control $u_t$, and $b(t,\tilde{x}_t)$ is a covariation function, describing its diffusion component in (2.1); $E_{s,x}$ is a conditional to the initial states ($s,x$) mathematical expectation, taken along the $\tilde{x}_t = \tilde{x}(t)$ trajectories.

Entropy functional (2.6) (as well as (1.1a)) is an *information indicator* of a *distinction* between the processes $\tilde{x}_t^a$ and $\tilde{x}_t^p$ by these processes' measures; it measures a *quantity of information* of process $\tilde{x}_t^p$ regarding process $\tilde{x}_t^a$. For the process' equivalent measures, this quantity is zero, and it takes a positive value for the process' nonequivalent measures.

The definition (1.1a,1.3) and (2.6) specifies *Radon-Nikodym's density* measure for a probability density measure, applied to entropy of a random process (Stratonovich 1973).

The *quantity of information* (2.6), (1.2a) is an equivalent of Kullback–Leibler's divergence (KL) for a continiuos random variable variables (Kullback 1959):

$$D_{KL}(P_{s,x}^a \parallel P_{s,x}^p) = \int_X \ln(\frac{dP_{s,x}^a}{dP_{s,x}^p}) dP_{s,x}^p = E_x[\ln \frac{P_{s,x}^a(d\omega)}{P_{s,x}^p(d\omega)}] = E_x[\ln p(\omega)] = -S_{ap}, \quad (2.7)$$

where $dP_{s,x}^a / dP_{s,x}^p$ is Radon–Nikodym derivative of probability $P_{s,x}^a(d\omega)$ with respect to probability $P_{s,x}^p(d\omega)$, and (2.7) determines a nonsymmetrical distance's measure between entropies $S_a$ and $S_p$ related to probabilities

$$P^a(X) = \int_X P_{s,x}^a(dx), \; P^p(X) = \int_X P_{s,x}^p(dx) \text{ accordingly.}$$

The KL measure is connected to both Shannon's conditional information and Bayesian inference (Jaynes 1998) of testing a priory hypothesis (probability distribution) by a posteriori observed probability distribution.

Finally, definition of information integral *information measure of transformation,* applied to a *process'* probabilities, generalizes some other information measures.

## 3. The information evaluation of a Markov diffusion process by an entropy functional measure on the process' trajectories

Advantage of the EF over Shannon's information measure consists in evaluating the inner connection and dependencies of the *random* process' states, produced at the generation of the process, which allows to measure the concealed information.

Such a functional information measure is able to accumulates the process' information, *hidden* between the information states, and, hence, brings more information then a sum of the Shannon's entropies counted for all process' separated states. We introduce a method of cutting off the process on the separated states by applying an impulse control, which is aimed to show that cutting off the EF integral information measure on the separated states' measures decreases the quantity of process information by the amount which was concealed in the connections between the separate states.

The $\delta$-cutoff of the diffusion process, considered below (Sec.3a), allows us to evaluate the *quantity* of information which the functional EF conceals, while this functional binds the correlations between the non-cut process states.

The cutoff leads to dissolving the correlation between the process cut-off points, losing the functional connections at these discrete points.

**3a.** *The step-wise and impulse controls' actions on functional* **(2.6)** *of diffusion process* $\tilde{x}_t$.

The considered control $u_t$ is defined as a piece-wise continuous function of $t \in \Delta$ having opposite stepwise actions:

$$u_+ \stackrel{def}{=} \lim_{t \to \tau_k + o} u(t, \tilde{x}_{\tau_k}), u_- \stackrel{def}{=} \lim_{t \to \tau_k - o} u(t, \tilde{x}_{\tau_k}), \qquad (3.1)$$

which is differentiable, excluding a set

$$\Delta^o = \Delta \setminus \{\tau_k\}_{k=1}^m, k = 1,...,m. \qquad (3.2)$$

The jump of the control function $u_-$ in (3.1) from a moment $\tau_{k-o}$ to $\tau_k$, acting on a *diffusion* process $\tilde{x}_t = \tilde{x}(t)$, might cut off this process after moment $\tau_{k-o}$. The cutoff diffusion process has the same drift vector and the diffusion matrix as the initial diffusion process. The entropy functional (2.5), defined through Radon-Nikodym's density measure (1.1), holds the properties of the cutoff controllable process.

The jump of the control function $u_+$ (3.1) from $\tau_k$ to $\tau_{k+o}$ might cut off the diffusion process *after* moment $\tau_k$ with the related additive functional in the form ( Prochorov, Rozanov 1973):

$$\varphi_s^{t-} = \begin{cases} 0, t \leq \tau_{k-o}; \\ \infty, t > \tau_k. \end{cases} \qquad (3.3)$$

The jump of the control function $u_+$ (3.1) from $\tau_k$ to $\tau_{k+o}$ might cut off the diffusion process *after* moment $\tau_k$ with the related additive functional

$$\varphi_s^{t+} = \begin{cases} \infty, t > \tau_k; \\ 0, t \leq \tau_{k+o}. \end{cases} \qquad (3.4)$$

At the moment $\tau_k$, between the jump of control $u_-$ and the jump of control $u_+$, we consider a control *impulse*

$$\delta u_{\tau_k}^{\mp} = u_-(\tau_{k-o}) + u_+(\tau_{k+o}). \qquad (3.5)$$

The related additive functional at a vicinity of $t = \tau_k$ acquires the form of an *impulse function*

$$\varphi_s^{t-} + \varphi_s^{t+} = \delta \varphi_s^{\mp}. \qquad (3.6)$$

summarizing (3.3) and (3.4).

The entropy functional at the localities of the control's switching moments (3.2) takes the values

$$S_- = E[\varphi_s^{t-}] = \begin{cases} 0, t \leq \tau_{k-o}; \\ \infty, t > \tau_k. \end{cases} \text{ and } S_+ = E[\varphi_s^{t+}] = \begin{cases} \infty, t > \tau_k; \\ 0, t \leq \tau_{k+o}. \end{cases} \qquad (3.7)$$

changing from 0 to $\infty$ and back from $\infty$ to 0 and acquiring an *absolute maximum* at $t > \tau_k$, between $\tau_{k-o}$ and $\tau_{k+o}$.

The related multiplicative functionals (2.3) are

$$p_s^{t-} = \begin{cases} 0, t \leq \tau_{k-o} \\ 1, t > \tau_k \end{cases} \text{ and } p_s^{t+} = \begin{cases} 1, t > \tau_k \\ 0, t \leq \tau_{k+o} \end{cases}, \qquad (3.7a)$$

which, according to (2.3), determine probabilities $P_{s,x}^a(d\omega) = 0$ at $t \leq \tau_{k-o}, t \leq \tau_{k+o}$ and $P_{s,x}^a(d\omega) = P_{s,x}^p(d\omega)$ at $t > \tau_k$.

For the cutoff diffusion process, the transitional probability (at $t \leq \tau_{k-o}$, $t \leq \tau_{k+o}$) turns to zero, the states $\tilde{x}(\tau-o), \tilde{x}(\tau+o)$ become independent, whereas their mutual time *correlations are dissolved*:

$$r_{\tau-o,\tau+o} = E[\tilde{x}(\tau-o)\tilde{x}(\tau+o)] \to 0. \tag{3.7b}$$

Entropy $\delta S_-^+(\tau_k)$ of additive functional $\delta\varphi_s^{\mp}$ (3.6), which is produced within, or at a border of the control impulse (3.5), is define by the equality

$$E[\varphi_s^{t-} + \varphi_s^{t+}] = E[\delta\varphi_s^{\mp}] = \int_{\tau_{k-o}}^{\tau_{k+o}} \delta\varphi_s^{\mp} P_\delta(d\omega), \tag{3.8}$$

where $P_\delta(d\omega)$ is a probability evaluation of impulse $\delta\varphi_s^{\mp}$.

Taking integral of the $\delta$-function $\delta\varphi_s^{\mp}$ between the above time intervals, we get at the border:

$$E[\delta\varphi_s^{\mp}] = 1/2 P_\delta(\tau_k) \text{ at } \tau_k = \tau_{k-o}, \text{ or } \tau_k = \tau_{k+o}. \tag{3.8a}$$

The impulse, produced by deterministic controls (3.5) for each process dimension $i = 1,...,n$, is a non random with

$$P_\delta^i(\tau_k) = 1, k = 1,...,m. \tag{3.8b}$$

This probability evaluates an average jump-diffusion transition probability (2.3), which is conserved during the jump. From (3.8)-(3.8b) we get the EF estimation at $t = \tau_k$ for each $i, k$ in the form

$$S_{\tau_k}^{\delta u_i} = E[\delta\varphi_{si}^{\mp}]_{\tau_k} = 1/2. \tag{3.9}$$

This entropy increment evaluates an information contribution at a vicinity of the discrete moments, when the impulse controls are applied. Since that, each information contribution

$$E[\varphi_{si}^{t-}]_{\tau_k} = S_{\tau_k}^{u_{i-}} \text{ and } E[\varphi_{si}^{t+}]_{\tau_k} = S_{\tau_k}^{u_{i+}} \tag{3.9a}$$

at a vicinity of $t = \tau_k$, produced by each of the impulse control's step-up and step-down function in (3.1), (3.5) accordingly, can be estimated by

$$S_{\tau_k}^{u_{i-}} = 1/4, u_- = u_-(\tau_k), \tau_{k-o} \to \tau_k; \text{ and } S_{\tau_k}^{u_{i+}} = 1/4, u_+ = u_+(\tau_k), \tau_k \to \tau_{k+o}, \tag{3.10}$$

where the entropy, according to its definition (1.1), is measured in the units of Nat (1 Nat $\cong$ 1.44bits).

Estimations (3.9), (3.10) determine the entropy functional's cutoff values at the above time's borders under actions of these controls, which decreases the quantity of the functional's information by the amount that had been concealed before the cutting the process correlations (3.7b).

The entropy functional (2.5), defined through the *Radon-Nikodym's probability density* measure (1.1), (3.7a), *holds* all properties of the considered cutoff controllable process.

The step-down control function $u_- = u_-(\tau_k)$ implements both transformation $\tilde{x}_t$ to $\varsigma_t$ and conversion of the entropy functional at $\tau_{k-o} \to \tau_k$ from its minimum at $t \leq \tau_{k-o}$ to the maximum at $\tau_{k-o} \to \tau_k$.

Therefore, maximization of the entropy functional under this control enables automatically transform $\tilde{x}_t$ to $\varsigma_t$, for which condition (2.1a) holds. The same way, the step-up control function $u_+ = u_+(\tau_k)$ implements both inverse transformation $\varsigma_t$ to $\tilde{x}_t$ and conversion the entropy functional at $\tau_k \to \tau_{k+o}$ from its maximum at $t > \tau_k$ $\tau_k \to \tau_{k+o}$ to the maximum at $\tau_k \to \tau_{k+o}$. Therefore, minimization of the entropy functional under this control enables automatically transform $\tilde{x}_t$ to $\varsigma_t$, for which condition (2.1a) also holds. Thus, impulse control (3.5) implements both minimax and maxmin transformations of the entropy functional (2.5), which is jumping from its *minimum to a maximum* and back.

The absolute maximum of the entropy functional at a vicinity of $t = \tau_k$ means that the impulse control delivers the *maximal amount* of information (3.9) from this transformation. This maximum, measured via the increment of additive functional, is limited by both the impulse's high and length. Whereas, the minimax is satisfied at *extracting* of random information, measured by EF under the impulse control actions.

In a multi-dimensional diffusion process, each of the step-up and step-down control, acting on the process' all dimensions, sequentially stops and starts the process. Dissolving the correlations, including those between all process' dimensions, leads to losing the correlation's connections at these discrete points.

The dissolved element of the correlation matrix at these moments provides independence of the cutting off fractions, evaluated by the Feller's kernel measure (Lerner 2012a). From that, it follows orthogonality of the correlation matrix for these cut off fractions. A sequence of the cutoff interventions in the process includes each time the impulse control's actions from the process' start up to its stopping for all dimensions.

### 3b. The estimation of information, hidden by the interstates' connection of the diffusion process.

How much information is lost at these points? The evaluated information effect of losing the functional's bound information at these points holds the amount of 0.5 Nats (~0.772 bits) (according to (3.9)) at each cut-off in the form of standard $\delta$-function for $k = i, i = 1,...,n$. Then $n$ of a such cut-off loses information

$$I_c \cong 0.772n, n = 1, 2, 3, ..., . \qquad (3.11)$$

Thus, the process functional's information measure encloses $I_c$ bits more, compared to the information measure applied separately to each of $n$ states of this process. The same result is applicable to a comparative information evaluation of the divided and undivided portions of an information process, measured by their corresponding EF.

This means that an information process holds more information than any divided number of its parts, and the entropy functional measure of this process is also able to evaluate the quantity of information that *connects* these parts.

As a result, the additive principle for the information of a process, measured by the EF, is *violated*:

$$\Delta S[\tilde{x}_t]\big|_s^T \geq \Delta S_1[\tilde{x}_t]\big|_s^{t_1} + \Delta S_2[\tilde{x}_t]\big|_{t_1+o}^{t_2} + \Delta S_m[\tilde{x}_t]\big|_{t_2+o}^{t_m} + \Delta S_T[\tilde{x}_t]\big|_{t_{m+o}}^T, ..., \qquad (3.11a)$$

where the process $\tilde{x}_t$ is cutting-off at the moments $t_1, t_1 + o; ....t_m, t_{m+o}; ....$

Therefore integral functional's measure accumulates more process information than the sum of the information measures in its separated states.

These entropy's increments evaluate the information contributions from the impulse controls at a vicinity of the above discrete moments, which are produced by both left stepwise control $u_-$ acting down, and the right stepwise control $u_+$ acting up. Together both of them extract the entropy functional's cutoff values, which decrease the quantity of the process' functional information by the amount that had been concealed before cutting the process correlations.

### 4. An optimal information transformation, its information measure and a minimax principle

Let us have information measure (2.6) for diffusion process (2.1), and *find* the condition of its optimization in the form

$$\min S[\tilde{x}_t]\big|_s^T = S[\tilde{x}_t^o]\big|_s^T, \qquad (4.1)$$

where is $\tilde{x}_t^o$ an extremal trajectory minimizing this functional, and $S[\tilde{x}_t^o]\big|_s^T$ is a functional minimizing $S[\tilde{x}_t]\big|_s^T$.

<u>Proposition 4.1.</u> Minimization of information transformation $E_{s,x}[-\ln p(\tilde{x}_t)]$ with $p = P_{s,x}^a / P_{s,x}^p$, where $P_{s,x}^a$ is differential probability of a priory process $\tilde{x}_t^a$, transformed to differential probability $P_{s,x}^p$ of *a posteriori* process $\tilde{x}_t^p$, leads to *converting* the probability measure $p$ in optimal $p_o = P_{s,x}^{ap} / P_{s,x}^{pd}$, which measures the transformation of random a priori trajectories $\tilde{x}_t^{ap}$ (with differential probability $P_{s,x}^{ap}$) to an *optimal predictable information process* $\tilde{x}_t^o \to x_t^{pd}$ (with probability measure $P_{s,x}^p \to 1$). (Integration of each differential probability determines the corresponding transitional probability distributions $P(s,x,t,B)$ according to (2.2)).

*Proof.* At $S[\tilde{x}_t]|_s^T = S_{ap}$, we get $\min S_{ap} = \min - E_{s,\tilde{x}}[\ln(P_{s,x}^a / P_{s,x}^{pd})] = \min(S_a - S_p)$,

where $S_a = E_{s,\tilde{x}}[-\ln(P_{s,x}^a)], S_p = E_{s,\tilde{x}}[-\ln(P_{s,x}^p)]$ are the entropy functionals of related processes.

At $\min S_{ap} = S_{ap}^o = (S_{ao} - S_{pd})$ we have

$$(S_{ao} - S_{pd}) = \min(S_a - S_p) , \qquad (4.2a)$$

where *a posteriori dynamic* process $x_t^{pd}$ with a differential probability $P_{s,x}^{pd} \to 1$ is a *predictable* process.

For such process

$$S_{pd} = -E_{s,x}[\ln P_{s,x}^{pd}] = -\ln P_{s,x}^{pd} = -\ln(P_{s,x}^{pd} \to 1) \to 0, \qquad (4.3)$$

and (4.2a) takes the form

$$\min(S_a - S_p) = \min S_a - \min S_p = \min S_a = \min S_{ao} = S_{ap}^o, \qquad (4.3a)$$

*satisfying to a minimum for information* transformation (4.1). This means, such *a posteriori dynamic* process $x_t^{pd}$ is an *optimal process* minimizing (4.1) as its *extremal*, and, by following (4.3), $x_t^{pd}$ is the *optimal predictable* process. •

<u>Comment 4.1.</u> (1). A random priori process $\tilde{x}_t$, holding an entropy functional measure (4.1): $S_{ap}[\tilde{x}_t]|_s^T$, at satisfaction of (4.3a), is measured by the optimal measure of a priori process in the form:

$$S_{ao}[\tilde{x}_t^{ap}] = \min S_{ap}[\tilde{x}_t] = S_{ap}^o. \qquad (4.4).$$

(2). The minimum in (4.2) not necessarily leads to (4.3a): all other transformed processes with $\min S_p > 0$, have

$$\min(S_a - S_p) < S_{ap}^o. \qquad (4.4a) •$$

<u>Corollary 4.1.</u>
Optimal transformation (4.1), which depends only on the entropy *minimum* for a priori process (4.4), has a *maximum of minimal entropy* $S_{ap}$ (4.1) among other transformed processes, therefore, it satisfies the *entropy's minimax principle*:

$$\max \min(S_a - S_p) = \min S_{ap} = S_{ap}^o, \qquad (4.5)$$

which we write in information form, at $I_{ap} = S_{ap}, \max P_{s,x}^p \to 1$:

$$\max \min - E_{s,x}[\ln(P_{s,x}^a / P_{s,x}^p)] = \max \min E_{s,x}[-\ln p(\omega)]] = \max \min I_{ap}. \qquad (4.6)$$

Hence, the sought dynamic process will be reached through such a minimal entropy transformation, which passes the minimum entropy of a priory process to a posteriori process with a maximum probability.

This means, the maximum of minimal information provides such information transformation a random process to a dynamic process, which approximates the random process with a maximal probability.

Otherwise: the *MiniMax information principle,* applied to a random process, implements optimal transformation (4.1) through an extraction of the process' information regularities with a maximal probability. •

Process $\tilde{x}_t$ gets dynamic properties, by holding a nonrandom functional $S[\tilde{x}_t]|_s^T$, which we call *information path functional* (IPF) on trajectories of its extremals $\tilde{x}_t$. The IPF information measure on extremals $\tilde{x}_t^o \to x_t^{pd}$ approximates random trajectories *of a priori process* $\tilde{x}_t \to x_t^{ap}$ with a *maximal* probability functional measure.

In this optimal approximation, a posteriori dynamic process brings a macroscopic evaluation of the priory random process, defining its *macrodynamic* process.

The IPF is a dynamic analogy of the EF, and the dynamic trajectories $x_t$ model diffusion process $\tilde{x}_t$, while trajectories $x_t$ integrate all $\tilde{x}_t$ cutoff correlations, being transformed by the controls.

The known *Maximum Entropy* (Information) principle, applied directly to a *random process,* would bring a related *MaxEnt* optimal process, having a *minimal probability distribution* (among all non *MaxEnt* optimal process), thereafter being the most *uncertain to indentify.*

While maximum entropy is associated with disordering and a complexity, minimum entropy means ordering and simplicity. That is why extracting a maximum of minimal information means also obtaining this maximum from a most ordered transformation.

Our goal of uncovering the regularity of a random process led us to transforming this process to a dynamic process, which holds these regularities and exposes them, whereas a randomness and uncertainty of the initial process could cover its regularities.

To retain the regularities of the transformed process, such transformation should not lose them, having a *minimal* entropy of the random process. The optimal transformation spends a minimal entropy (of a lost) for transforming each priory random process to its dynamic model.

Thus, the transformed process that holds regularities should satisfy some *variation principle*, which according to R. Feynman 1963, could be applied to a process as a mathematical form of a law expressing the process regularities.

We formulate this law through imposing the information *MiniMax* variation principle (VP) on the random process, which brings both its optimal transformation, minimizing the entropy for a random dynamic process and maximizes this minimum.

Solving the VP allows us finding the IPF functional $S[\tilde{x}_t]|_s^T$ and its extremals $x_t$, which approximate a posteriori random process with a maximum probability on its dynamic trajectories.

Thus, *MiniMax* principle is implemented by the VP through minimization of entropy functional (EF) of random process, whose minimum is maximized by information path functional (IPF) of information macrodynamics.

The EF-IPF approach converts the *uncertainty* of a random process into the *certainty* of a dynamic information process.

## 5. Solution of the variation problem

Applying a variation principle (4.1) to the entropy functional, we consider an integral functional

$$S = \int_s^T L(t,x,\dot{x})dt = S[x_t] \;, \tag{5.1}$$

which minimizes the entropy functional (2.6) of the controlled process in the form

$$\min_{u_t \in KC(\Delta,U)} \tilde{S}[\tilde{x}_t(u)] = S[x_t], \; Q \in KC(\Delta, R^n) \,. \tag{5.1a}$$

*Proposition 5.1.*

(a). An *extremal solution* of variation problem (5.1a,5.1) for the entropy functional (3.4) brings the following equations of extremals for a vector $x$ and a conjugate vector $X$ accordingly:

$$\dot{x} = a^u, \; (t,x) \in Q \,, \tag{5.2}$$

$$\dot{X} = -\partial P/\partial x - \partial V/\partial x \,, \tag{5.3}$$

where $P = (a^u)^T \dfrac{\partial S}{\partial x} + b^T \dfrac{\partial^2 S}{\partial x^2} \,,$ (5.4)

$S(t,x)$ is function of action on extremals (5.2,5.3); $V(t,x)$ is the function (3.3) for the additive functional (1.6), in (1.3) and in (3.2), which defines the probability function $\tilde{u} = \tilde{u}(t,x)$; $P(t,x)$ is a potential function on the extremals.

*Proof.* Using the Jacobi-Hamilton (JH) equations for function of action $S = S(t,x)$ (Gelfand, Fomin 1962), defined on the extremals $x_t = x(t), (t,x) \in Q$ of functional (5.1), we have

$$-\frac{\partial S}{\partial t} = H, H = \dot{x}^T X - L, \tag{5.5}$$

where $X$ is a conjugate vector for $x$ and $H$ is a Hamiltonian for this functional.
(All derivations here and below have vector form).
Let us consider the distribution of functional (2.6) on $(t, x) \in Q$ as a function of current variables $\tilde{S} = \tilde{S}(t, x)$, which satisfies the Kolmogorov (K) equation (Gichman and Scorochod, 1973, Dynkin,1960, others), applied to the math expectation of functional (2.6) in the form:

$$-\frac{\partial \tilde{S}}{\partial t} = (a^u)^T \frac{\partial \tilde{S}}{\partial x} + b \frac{\partial^2 \tilde{S}}{\partial x^2} + 1/2(a^u)^T (2b)^{-1} a^u . \tag{5.5a}$$

From condition (5.1a) it follows

$$\frac{\partial S}{\partial t} = \frac{\partial \tilde{S}}{\partial t}, \frac{\partial \tilde{S}}{\partial x} = \frac{\partial S}{\partial x}, \tag{5.5b}$$

where for the JH we have $\frac{\partial S}{\partial x} = X, -\frac{\partial S}{\partial t} = H$.

This allows us to join Eqs (5.5), (5.5a) and (5.5b) in the form

$$-\frac{\partial \tilde{S}}{\partial t} = (a^u)^T X + b \frac{\partial X}{\partial x} + 1/2 a^u (2b)^{-1} a^u = -\frac{\partial S}{\partial t} = H, \tag{5.6}$$

where a dynamic Hamiltonian is represented the form $H = V + P$, which includes function $V(t, x)$ and function

$$P(t, x) = (a^u)^T X + b^T \frac{\partial X}{\partial x}. \tag{5.7}$$

Applying to (5.6) Hamilton equation $\frac{\partial H}{\partial X} = \dot{x}$ and $\frac{\partial H}{\partial x} = -\dot{X}$, we get the extremals for the vectors $x$ and $X$ in the forms (5.2) and (5.3) accordingly. •

(b). A *minimal solution* of variation problem (5.1a, 5.1) for the entropy functional (3.4) brings the following equations of extremals for $x$ and $X$ accordingly:

$$\dot{x} = 2b X_o, \tag{5.8}$$

$$\dot{X}_o = -2 H_o X_o, \tag{5.9}$$

satisfying condition $\min_{x(t)|_{t \to \tau}} P = P[x(\tau)] = 0$. \hfill (5.10)

Condition (5.10) is a dynamic constraint, imposed on the solutions (5.2), (5.3) at some set of the functional's field $Q \in KC(\Delta, R^n)$, where the following relations hold:

$$Q^o \subset Q, Q^o = R^n \times \Delta^o, \Delta^o = [0, \tau], \tau = \{\tau_k\}, k = 1,...,m \text{ for process } x(t)_{t=\tau} = x(\tau). \tag{5.11}$$

This condition, appled to function $\tilde{S} = \tilde{S}(t, x)$ at the moment (5.11), brings the constraint in the form

$$E_{x,\tau}[(a^u)^T \frac{\partial \tilde{S}}{\partial x} + b \frac{\partial^2 \tilde{S}}{\partial x^2}] = 0, a^u = a^u(t, x), b = b(t, x). \tag{5.11a}$$

(c). Hamiltonian in (5.9) $H_o = -\frac{\partial S_o}{\partial t}$ \hfill (5.12)

is defined for the function of action $S_o(t, x)$, which on the extremals (5.8, 5.9) satisfies the condition $\min(-\partial \tilde{S} / \partial t) = -\partial \tilde{S}_o / \partial t$. \hfill (5.13)

Hamiltonian (5.6) and Eq. (5.8) determine a second order differential Eq. of the extremals:

$$\ddot{x} = \dot{x}[\dot{b}b^{-1} - 2H]. \tag{5.14}$$

*Proof.* Using (5.4) and (5.6), we find the equation for Lagrangian in (5.1) in the form

$$L = -b \frac{\partial X}{\partial x} - 1/2 \dot{x}^T (2b)^{-1} \dot{x} .\qquad(5.15)$$

Since on extremals $x_t = x(t)$ (5.2,5.3), both $a^u$ and $b$ (in 1.1, 1.6) are nonrandom, after their substitution to (5.1) we get the integral functional $\tilde{S}$ on the extremals:

$$\tilde{S}[x(t)] = \int_s^T 1/2 (a^u)^T (2b)^{-1} a^u dt ,\qquad(5.15a)$$

which should satisfy the variation conditions (5.1a), or

$$\tilde{S}[x(t)] = S_o[x(t)],\qquad(5.15b)$$

where both integrals are determined on the same extremals.
From (5.15), (5.15a,b) it follows

$$L_o = 1/2 (a^u)^T (2b)^{-1} a^u, \text{ or } L_o = \dot{x}^T (2b)^{-1} \dot{x} .\qquad(5.16)$$

Both expressions for Lagrangian (5.15) and (5.16) coincide on some extremals, where potential (5.10) takes form

$$P_o = P[x(t)] = (a^u)^T (2b)^{-1} a^u + b^T \frac{\partial X_o}{\partial x} = 0 ,\qquad(5.17)$$

for Hamiltonian (5.12) and the function of action $S_o(t,x)$ satisfying (5.13). From (5.15b) it also follows

$$E\{\tilde{S}[x(t)]\} = \tilde{S}[x(t)] = S_o[x(t)] .\qquad(5.17a)$$

At this condition, after substituting relation (5.17) to (5.5a) we come to (5.11a), which is satisfied at the discrete moments (5.11).
Applying to (5.16) the Lagrange's equation

$$\frac{\partial L_o}{\partial \dot{x}} = X_o,\qquad(5.17b)$$

we get

$X_o = (2b)^{-1} \dot{x}$ (5.17c) and extremals (5.8).

Lagrangian (5.16) satisfies the principle maximum (Alekseev, Tichomirov, Fomin 1979) for functional (5.15a); from the principle it also follows (5.17a). Thus, functional (5.1) reaches its minimum on extremals (5.8), while on the extremals (5.2), (5.3) this functional reaches some extremal values corresponding to Hamiltonian (5.6).
This Hamiltonian, at satisfaction of (5.17), reaches its minimum:

$$\min H = \min[V + P] = 1/2 (a^u)^T (2b)^{-1} a^u = H_o ,\qquad(5.18)$$

from which it follows

$$V = H_o \qquad(5.19a) \qquad\qquad \text{at } \min_{x(t)|_{t \to \tau}} P = P[x(\tau)] = 0 .\qquad(5.19b)$$

Function $(-\partial \tilde{S}(t,x)/\partial t) = H$ in (5.6) on extremals (5.2,5.3) reaches a *maximum* when the constraint (5.10) *is not* imposed. Both the minimum and maximum are conditional with respect to the constraint imposition.
The variation conditions (5.18), imposing constraint (5.10) at discrete moments $\tau = \{\tau_k\}$ (5.11), selects Hamiltonian

$$H_o = -\frac{\partial S_o}{\partial t} = 1/2 (a^u)^T (2b)^{-1} a^u \qquad(5.20)$$

on the extremals (5.8,5.9).
The variation principle identifies two Hamiltonians: $H$ satisfying (5.6) with function of action $S(t,x)$, and $H_o$ (5.20), whose function action $S_o(t,x)$ reaches absolute minimum and coincides with $S(t,x)$ at the moments (5.11) of imposing constraint (5.11a).
By substituting (5.2) and (5.17b) in both (5.16) and (5.20), we get the Lagrangian and Hamiltonian on the extremals:

$$L_o(x, X_o) = 1/2 \dot{x}^T X_o = H_o .\qquad(5.21)$$

Using $\dot{X}_o = -\partial H_o / \partial x$, we have $\dot{X}_o = -\partial H_o / \partial x = -1/2\dot{x}^T \partial X_o / \partial x$, and from constraint (5.10), we get
$\partial X_o / \partial x = -b^{-1}\dot{x}^T X_o$, and $\partial H_o / \partial x = 1/2\dot{x}^T b^{-1}\dot{x}^T X_o = 2H_o X_o$,
which after substituting (5.17b) leads to extremals (5.9).
By differentiating (5.8) we get a second order differenti al Eqs on the extremals:
$\ddot{x} = 2b\dot{X}_o + 2\dot{b}X_o$, which after substituting (5.9) leads to
$\ddot{x} = 2X_o[\dot{b} - 2bH]$, or to (5.14). •

Corollary 5.1. At imposing the constraint (5.17) on (5.6), we get the variation condition for the constraint integral form
$S_c = S_c[x(t)]$: $\frac{\partial S_c}{\partial t} = b\frac{\partial X_o}{\partial x} = -(a^u)^T X_o$,

which is connected to Hamiltonian (5.20) by relation $\frac{\partial S_c}{\partial t} = 2H_o$.

Function (5.6) on extremals (5.8-5.9) is expressed by this Hamiltonian in the form $\frac{\partial S}{\partial t} = 3H_o$.

Then we get relation $\frac{\partial S_c}{\partial t} = 3/2 \frac{\partial S}{\partial t}$, which is satisfied at the moments (5.11). Since functional (5.1) in the form

$S = \int_{t_k} Hdt$ is invariant on the extremals (5.2-5.3) during each interval $t_k$ of the extremal movement, its increment

$\Delta S_c[x(t_k)] = inv$, $k = 1,...,m$ \hfill (5.21a)

is preserved between the punched points (5.11). The constraint's form (5.21a) expresses the VP information invariant that depends on both functions $a^u, b$ in (2.6) and (5.4). Within the interval of imposing constraint (5.7) (before the constraint (5.5) reach its minimum at the moments (5.19a)), the diffusion processes with entropy functional (2.6) enables generate the process' dynamics, satisfying variation Eqs (5.1-5.3). While holding the constraint (5.19a) during a time interval $\Delta t_k$ of completion (5.18) provides the process dynamics, satisfying Eqs (5.8-5.9). The invariant relation (5.21a) will be used to find both local information invariants (Secs.7, 8) and interval of applying control $t_k$. •

Corollary 5.2.
The control action on equation (5.6), which implements the variation conditions (5.1a) at the set of discrete moments (5.11), requires *turning* the constraint (5.4) *on* with changing the process' function $-\partial \tilde{S} / \partial t$ from its *maximum* to a *minimum*.
(1).While both Hamilton Eqs (with Hamilatonians in forms (5.6) and (5.20)) accordingly) provide *extremal* solutions *minimizing* functional (5.1), the extremal with Hamiltonian (5.6) minimizes this functional with a *maximal* speed $|\partial \tilde{S} / \partial t| > |\partial \tilde{S}_o / \partial t|$ compared to the extremal with Hamiltonian (5.20).
(2).These maximal extremal solutions approximate a priori diffusion process with a minimum of maximal probability, satisfying $\max |\partial \tilde{S} / \partial t|$, and a maximum of the probability, satisfying $\min |\partial \tilde{S} / \partial t| = |\partial \tilde{S}_o / \partial t|$ accordingly. •

Solution of this variation problem's (VP) (Lerner 2007) *automatically brings the constraint*, imposing *discretely* at the states' set (5.4a) by the applied optimal controls (synthesized in Sec.6), which change the entropy derivation from its maximum to its minimum.

Consequently, an extremal is determined by imposing such dynamic constraint during the control actions, which select the extremal *segment* from the initial random process (Fig.1a). •

Below we find the limitations on completion of constraint equation (5.4), which also restrict the controls action, specifically when it should be turn off.

*Proposition 5.2.* Let us consider diffusion process $\tilde{x}(s,t)$ at a locality of states $x(\tau_k - o), x(\tau_k), x(\tau_k + o)$, formed by the impulse control's cut-off action (Sec3a), where the process is cutting off *after* each moment $t \leq \tau_k - o$ -at $t > \tau_k$, and each moment $t \geq \tau_k + o$ is *following* to the cut-off, with $(\tau_k - o) < \tau_k < (\tau_k + o)$.

The additive and multiplicative functionals (Sec.3a) satisfy Eqs (3.7,3.7a) at these moments.

Then the constraint (5.11a) acquires the form of operator $\tilde{\tilde{L}}$ in Eq.

$$-\frac{\partial \Delta \tilde{S}}{\partial s} = \tilde{\tilde{L}} \Delta \tilde{S}, \Delta \tilde{S}(s,t) = \begin{cases} 0, t \leq \tau_k - o; \\ \infty, t > \tau_k; \end{cases} \quad (5.22)$$

which at $\Delta \tilde{S}(s, t \leq \tau_k - o) = 0$ (5.22a) satisfies Eq

$$\tilde{\tilde{L}} = (a^u)^T \frac{\partial}{\partial x} + b \frac{\partial^2}{\partial x^2} = 0. \quad (5.23)$$

The *proof* follows from (Prokhorov, Rosanov 1973), where it is shown that $\Delta \tilde{S}(s, t \leq \tau_k - o) = E_{s,x}[\varphi_s^{t-}]$ satisfies to the operator $\tilde{\tilde{L}}$ in Eq (5.22), which is connected with operator $\tilde{L}$ of the initial K Eqs (5.5a) by relation $\tilde{\tilde{L}} = \tilde{L} - 1/2(a^u)^T (2b)^{-1} a^u$.

From these relations, at completion of (5.22a), we get (5.22b) and then

$$E_{s,x}[(a^u)^T \frac{\partial \Delta \tilde{S}}{\partial x} + b \frac{\partial^2 \Delta \tilde{S}}{\partial x^2}] = 0, \quad (5.24)$$

where $\Delta \tilde{S}(s, t \leq \tau_k - o) = E_{s,x}[\varphi_s^{t-}] = S_-$ is the process' functional, taken before the moment of cutting off, when constraint (5.10, 5.11a) is still imposed. •

From the same reference it also follows that solutions of (5.24) allow classifying the states $x(\tau) = \{x(\tau_k)\}, k = 1,...,m$, considered to be the *boundary* points of a diffusion process at $\lim_{t \to \tau} \tilde{x}(t) = x(\tau)$.

A boundary point $x_\tau = x(\tau)$ *attracts* only if the function

$$R(x) = \exp\{-\int_{x_o}^{x} a^u(y) b^{-1}(y) dy\}, \quad (5.25)$$

defining the general solutions of (5.25), is integrable at a locality of $x = x_\tau$, satisfying the condition

$$|\int_{x_o}^{x_\tau} R(x) dx| < \infty. \quad (5.25a)$$

A boundary point *repels* if (5.25a) does not have the limited solutions at this locality; it means that the above Eq. (5.25) is not integrable in this locality. • The boundary dynamic states carry *hidden dynamic* connections between the process' states.

Comments 5.1.

(1). The constraint, imposed at $t \leq (\tau_k - o)$, corresponds to the VP action *on the macrolevel*, whereas this action produces the cut-off at the following moment $\tau_k > (\tau_k - o)$ on the *microlevel* (both random and quantum (Lerner 2010b, 2012b), when the constraints is turning off. Hence, the moment of imposing constraint (5.10) on the dynamic macrolevel coincides with the moment of imposing constraint (5.23) on the microlevel, and the same holds true for the coinciding moments of turning both constraints off, which happen simultaneously with applying step-down control $u_-(\tau_k - o)$ (as a left part of the impulse control (3.5) (Fig. 1b) that cuts-off the random process).

The state $x(\tau_k - o)$, prior to the cuttingoff, has a minimal entropy $S[x(\tau_k - o)] \to 0$, which with a maximal probability puts it close to the entropy of a nonrandom dynamic macrostate on the extremal at the same moment of imposing the constraint. Under the $u_-(\tau_k - o)$ control action (Sec3a), the state $x(\tau_k - o)$ with a minimal entropy moves to a random state $x(\tau_k)$ with a maximal entropy, which turns the dynamics to a randomness.

(2).Under the $u_+(\tau_k + o)$ control action, state $x(\tau_k)$, with a maximal entropy $S[x(\tau_k)] \to \infty$ (according to (3.7)), moves to state $x(\tau_k + o)$, which, *after* the cutting-off, has a minimal entropy $S[x(\tau_k + o)] \to 0$ as it approaches dynamic state $x(\tau_{k+1}^1)$. This means that the state, currying a maximal information, transfers it to the state with a minimal information (and a maximal probability), which gets it with this maximal probability (Sec 4). The state $x(\tau_k + o)$, having a maximal probability, could be transferred to a dynamic process as its starting point (on a following extremal).

Moreover, this state absorbs an increment of the entropy functional Sec.3a, (3.9a),(3.10) at each moment $\tau_k + o : S[x(\tau_k + o)] \cong 1/4 Nats$, which is transported to a starting extremal segment with the macrostate $x(\tau_k + o)$, as its primary entropy contribution to the macrodynamics from control $u_+(\tau_k + o)$ (at each $k = i, i = 1,...n$).

The same way, the control $u_-(\tau_k - o)$ action had transfered the increment of the entropy functional (3.10) at the moment $\tau_k - o : S[x(\tau_k - o)] \cong 1/4 Nats$ from the ending point of a previous extremal segment to the diffusion process.

This means, both ending and starting points of each extremal's segment possess the same two parts of a total information contribution (3.9) getting it from the diffusion process.

This "hidden information", obtained from both bound *dynamic* connections between the process' states and the cutoff information contribution is a source of the information macrodynamics.

(3). A locality $x(\tau_k)$ of its border states $((x(\tau_k - o), x(\tau_k + o))$ forms *"punched"discrete points* (DP) $(..., x(\tau_{k-1}), x(\tau_k), x(\tau_{k+1}),...)$ of the space $Q^o = R^n \times \Delta^o$, which establish a link between the microlevel's diffusion and macrolevel's dynamics. The macroequations (5.2, 5.3) act along each extremal, only approaching these points to get information from the diffusion process with the aid of an optimal control.

The impulse control intervenes in a stochastic process (like that in (Yushkevich 1983)), extracts information (from the above locality) and transfers it to macrodynamics. The information contribution in this locality from the entropy functional (EF) coincides with that for the information path functional (IPF) on each extremal segment's ending and starting points.

The constraint equation is the *main mathematical structure* that distinguishes the equations of diffusion stochastics from related *dynamic* equations; at imposing the constraint, both equation's solutions coincide (at the process' border states). •

(4). The optimal control, which implements VP by imposing the constraint (in the form (5.23)) on the microlevel and (in the form (5.10)) on the macrolevel, *and* generating the macrolevel's dynamics, should have a *dual and simultaneous action* on both diffusion process and its dynamic model.

It's seen that the impulse control (IC), composed by two step-controls SP1, SP2, Fig.1b (forming the IC's left and right sides accordingly, Sec.3), which provides the cutoff of the diffusion process, can turn *on* and *off* the constraint on the microlevel, and in addition to that, transfers the state with maximal probability, produced after the cutoff, to start the macrolevel's dynamics with initial entropy contribution gaining from the cutoff.

The IC, performing the above dual operations, implements the VP, as its *optimal control*, which extracts most probable states from a random process, in the form of a $\delta$-probability distribution.

We *specify* the following control's functions:

– The IC, applied to diffusion process, provides a sharp maximum of the process information, which coincides with the information of starting at the same moment the model's dynamic process. That implements the principle of maximal entropy at this moment;

– The start of the model's dynamic process, is associated with imposing the dynamic constraint (DC) on the equation for the diffusion process with the SP2 dual action, which also keeps holding along the extremal;

– At the moment when the DC finishes imposing the constraint, the SP2 stops, while under its stepwise-down action (corresponding to control SP1), the dynamics process on the extremal meets the punched ("bounded") locality of diffusion process;

– At this locality, the IC is applied (with both SP1, stopping the DC, and the SP2, starting the DC at the next extremal segment), and all processes sequentially repeat themselves, concurrently with the time course of the diffusion process;

– The starting impulse control binds the process' maximum probability state, with the following start of the extremal movement, which keeps the maximum probability during its dynamic movement. •

(5). Because the dynamic process depends on the moments' of controls' actions, which also depend on the information getting from the random process, a probability on trajectories of the dynamic process is less than 1.

Such dynamic process is *a quasi-deterministic process,* it's a *Bernstein-Markov process* (Cruzeiro, Wu, Zambrini 98) described here in terms of information dynamics, which is generated through the intervention of deterministic step-up control's action in the Brownian movement (Lerner 2012b). •

(6). Since both the constraint imposition and the control action are limited by punched points $x(\tau_k), x(\tau_{k+1})$, the control, starting an extremal movement at the moment $\tau_k^o = \tau_k + o$, should be *turned off* at a moment $\tau_k^1 = \tau_{k+1} - o$ preceding to the following punched point $\tau_{k+1}$. And to continue the extremal movement, the control should be turned *on* again at the moment $\tau_{k+1}^o = \tau_{k+1} + o$ following $\tau_{k+1}$ to start a next extremal segment. This determines a *discrete* action of the stepwise control *during each interval of the extremal movement* $t_k = \tau_k^1 - \tau_k^o, t_{k+1} = \tau_{k+1}^1 - \tau_{k+1}^o, k = 1, ..., m$, where $m$ is the number of applied controls. Both the control's amplitude and time intervals will be found in Sec.6.

The constraint limits the time length and path of the extremal segment between the punched localities.

(A very first control is applied at the moment $\tau_o^o = s + o$, following the process Eqs' initial condition at $t = s$.)

The process continuation requires connecting the extremal segments between the punched localities by the joint stepwise control action: with a $k$- control SP1 *turning off* at the moment $\tau_k^1$ while transferring to locality $\tau_{k+1}$, and a next $k+1$- control SP2, which, while transferring from $\tau_{k+1}$ locality, is *turning on* the following extremal segment at moment $\tau_{k+1}^o$. These controls provides a *feedback* from a *random process to macroprocess* and then *back* to a random process.

By imposing constraints (5.10) and (5.23) during each time interval $t_k$, both controllable random process (as an object) and its *dynamic feedback model* become equal probable, being *prepared* (during this time delay) for getting new information and the optimal control action utilizing this information.

Both stepwise controls form an impulse IC control function $\delta(x(\tau_k^1), x(\tau_{k+1}), x(\tau_{k+1}^o))$ acting between moments $\tau_k^1, \tau_{k+1}, \tau_{k+1}^o$ (or $\tau_{k-1}^1, \tau_k, \tau_k^o$) and implementing relation (3.5), which brings the EF peculiarities (3.7) at these moments, making the impulse control (3.5) a part of the VP implementation.

Such a control imposes the constraint in the forms (5.10,5.23) on the initial random process, selecting its border points, as well as it starts and terminates the extremal movement between these points, which models each segment of the random process by the segment's dynamics. •

## 6. The optimal control synthesis, model identification, and specifics of both DC and IPF solutions

Writing equation of extremals $\dot{x} = a^u$ in a dynamic model's traditional form (Alekseev *et al* 1979):

$$\dot{x} = Ax + u, u = Av, \dot{x} = A(x + v),  \qquad (6.1)$$

where $v$ is a control reduced to the state vector $x$, we find optimal control $v$ that solves the initial variation problem (VP) and identifies matrix $A$ under this control's action.

*Proposition 6.1.*

The reduced control is formed by a feedback function of macrostates $x(\tau) = \{x(\tau_k)\}, k = 1, ..., m$:

$$v(\tau) = -2x(\tau), \qquad (6.2)$$

or
$$u(\tau) = -2Ax(\tau) = -2\dot{x}(\tau), \quad (6.2a)$$

at the DP localities of moments $\tau = (\tau_k)$ (5.11), and the matrix $A$ is identified by the equation

$$A(\tau) = -b(\tau)r_v^{-1}(\tau), r_v = E[(x+v)(x+v)^T], b = 1/2\dot{r}, r = E[\tilde{x}\tilde{x}^T] \quad (6.3)$$

through the above correlation functions, or directly, via the dispersion matrix $b$ from (2.1):

$$|A(\tau)| = b(\tau)(2\int_{\tau-o}^{\tau} b(t)dt)^{-1} > 0, \ \tau - o = (\tau_k - o), k = 1...,m. \quad \bullet \quad (6.3a)$$

*Proof.* Using Eq. for the conjugate vector (5.3), we write the constraint (5.10) in the form

$$\frac{\partial X}{\partial x}(\tau) = -2XX^T(\tau), \quad (6.4)$$

where for model (6.1) we have

$$X = (2b)^{-1}A(x+v), X^T = (x+v)^T A^T (2b)^{-1}, \frac{\partial X}{\partial x} = (2b)^{-1}A, \ b \neq 0, \quad (6.4a)$$

and (6.4) acquires the form

$$(2b)^{-1}A = -2E[(2b)^{-1}A(x+v)(x+v)^T A^T (2b)^{-1}], \quad (6.4b)$$

from which, at a nonrandom $A$ and $E[b] = b$, the identification equations (6.3) follow up.

The completion of both (6.3) and (6.4) is reached with the aid of the control's action, which we find using (6.4a) in the form

$$A(\tau)E[(x(\tau)+v(\tau))(x(\tau)+v(\tau))^T] = -E[\dot{x}(\tau)x(\tau)^T], \text{ at } \dot{r} = 2E[\dot{x}(\tau)x(\tau)^T]. \quad (6.5)$$

This relation after substituting (6.1) leads to

$$A(\tau)E[(x(\tau)+v(\tau))(x(\tau)+v(\tau))^T] = -A(\tau)E[(x(\tau)+v(\tau))x(\tau)^T], \quad (6.5a)$$

and then to

$$E[(x(\tau)+v(\tau))(x(\tau)+v(\tau))^T + (x(\tau)+v(\tau))x(\tau)^T] = 0, \quad (6.5b)$$

which is satisfied at applying the control (6.2). Control (6.2a) can be formed directly from (6.2a) by measuring $\dot{x}(\tau)$.

Since $x(\tau)$ is a discrete set of states, satisfying (5.11), (5.13), the control has a discrete form, applied at this set.

Each stepwise control (6.2) with its inverse amplitude $-2 x(\tau)$, doubling the controlled state $x(\tau)$, is applied at beginning of each extremal segment and acts during the segment's time interval. This control imposes the constraint (in the form (6.4), (6.4b)), which follows from the variation conditions (5.1a), and, therefore, it implements this condition.

The same control, applied to both random process and the extremal segment, transforms $\tilde{x}_t^a$ to $\tilde{x}_t^p = \tilde{x}_t^{ao}$, and then, by imposing the constraint, transforms $\tilde{x}_t^{ao}$ to $\tilde{x}_t^{pd}$ extremals. During the last transformation, this control also initiates the identification of matrix $A(\tau)$ following (6.3,6.3a). These two stepwise controls perform both transformations as a single impulse control which sequentially starts and terminates the constraint on each segment, while its jump down and up actions at each segment's punched locality extracts the hidden information.

Finding this control here just *simplifies* some results of Theorem (Lerner 2007). •

Corollary 6.1. The control, that turns the constraint on, creates a Hamilton dynamic model with complex conjugated eigenvalues of matrix $A$. After the constraint's termination (at the diffusion process' boundary point), the control transforms this matrix to its *real* form (on the boundary), which is identified by diffusion matrix in (6.3). •

*Proposition* 6.2.

Let us consider the controllable dynamics of a closed system, described by operator $A^v(t,\tau)$ with eigenfunctions $\lambda_i^v(t_i,\tau_k)_{i,k=1}^{n,m}$, whose matrix equation:

$$\dot{x}(t) = A^v x(t), \quad (6.6)$$

includes the feedback control (6.2) and leads to the same form of a drift vector for both models (6.1) and (6.6):
$$a^u(\tau, x(\tau,t)) = A(\tau,t)(x(\tau,t) + v(\tau)); A(\tau)(x(\tau) + v(\tau)) = A^v(\tau)x(\tau) \quad . \tag{6.6a}$$

Then the followings hold true:

(1)-Matrix $A(t,\tau)$ under control $v(\tau_k^o) = -2x(\tau_k^o)$, applied during time interval $t_k = \tau_k^1 - \tau_k^o : A(t_k, \tau_k^1)$ depends on initial matrix $A(\tau_k^o)$, taken at the moment $\tau_k^o$, by the Eq

$$A(t_k, \tau_k^1) = A(\tau_k^o)\exp(A(\tau_k^o)t_k)[2 - \exp(A(\tau_k^o)t_k)]^{-1} . \tag{6.6b}$$

(2)- The identification Eq.(6.3) at $\tau_k^1 = \tau$ gets form

$$A^v(\tau) = -A(\tau) = 1/2b(\tau)r_v^{-1}(\tau), b(\tau) = 1/2\dot{r}_v(\tau) , \tag{6.6c}$$

whose covariation function $r_v(\tau_k^o)$, starting at the moment $\tau_k^o$, by the end of this time interval acquires form

$$r_v(\tau_k^1) = [2 - \exp(A(\tau_k^o)t_k)]r(\tau_k^o)[2 - \exp(A(\tau_k^o)t_k)] . \tag{6.6d}$$

(3a)-At the moment $\tau_k^o + o$ following $\tau_k^o$ of applying control $v(\tau_k^o) = -2x(\tau_k^o)$, the controllable matrix gets form

$$A^v(\tau_k^1)_{t_k \to 0} = A^v(\tau_k^o + o) = -A(\tau_k^o), \tag{6.7}$$

changing the sign of the initial matrix.

(3b)-When that control, being applied at the moment $\tau_k^1$, ends the dynamic process on extremals in the following moment, at $x(\tau_k^1 + o) \to 0$, function $a^u = A^v x(t)$ in (6.6) turns to

$$a^u(x(\tau_k^1 + o)) \to 0; \tag{6.7b}$$

which brings (6.6a) to its dynamic form $a^u = A(\tau_k^1 + o)v(\tau_k^1 + o) \to 0$ that requires turning the control off, at $v(\tau_k^1 + o) \to 0$.

(3c). At (6.7b), stochastic Eq. (2.1) is defined only by its diffusion component, which according to (6.3) allows identifying dynamic matrix $A(\tau_k^1 + o)$, being transformed in the following moment $\tau_{k+1}^1 : A(\tau_k^1 + o) \to A(\tau_{k+1}^1)$, via correlation matrix $r(\tau_{k+1}^1)$ by relation $A(\tau_{k+1}^1) = 1/2\dot{r}(\tau_{k+1}^1)r^{-1}(\tau_{k+1}^1)$, where $r(\tau_{k+1}^1) = E[\tilde{x}(t)\tilde{x}(t + \tau_{k+1}^1)^T] = r^v(\tau_{k+1}^1)_{v(\tau_k^1 + o) \to 0}$.

*Proof* (1). Applying control $v(\tau_k^o) = -2x(\tau_k^o)$, which imposes the constraint on both (6.1) and (6.6) during time interval $t_k = \tau_k^1 - \tau_k^o$ (starting it at $\tau_k^o$ and terminating at $\tau_k^1$), we get the solutions of (6.1) by the *end* of this interval:

$$x(\tau_k^1) = x(\tau_k^o)[2 - \exp(A(\tau_k^o)t_k)]. \tag{6.7c}$$

Substituting the solution (6.7c) to the right side of $\dot{x}(\tau_k^1) = A^v(\tau_k^1)x(\tau_k^1)$ and the derivative of (6.7c) to the left side, we come to

$$-x(\tau_k^o)(A(\tau_k^o)t_k)\exp(A(\tau_k^o)t_k) = A^v(\tau_k^o)x(\tau_k^o)[2 - \exp(A(\tau_k^o)t_k)]),$$

or to connection of both matrixes $A^v(\tau_k^1)$ and $A(\tau_k^1)$ (at the interval end) with the matrix $A(\tau_k^o)$ (at the interval beginning) for the closed system (6.6) in the forms:

$$A^v(t_k, \tau_k^1) = -A(\tau_k^o)\exp(A(\tau_k^o)t_k)[2 - \exp(A(\tau_k^o)t_k)] , \tag{6.7d}$$

and $A^v(\tau_k^1) = -A(\tau_k^1)$ by the moment $\tau_k^1$, from which we get (6.6b).

*Proof* (2) (6.6c) follows from (6.7d) at $\tau = \tau_k^1$, and we get (6.6d) directly after substitution solution (6.7) to

$$r(\tau_k^1) = E[x(\tau_k^1)x(\tau_k^1)^T].$$

*Proof* (3a) follows from relation (6.6d).
*Proof* (3b) (6.6c) follows from relations

$$a^u(x(\tau^1_{k+1}), v(\tau^1_{k+1})) = Ax(\tau^1_{k+1}) + A(-2x(\tau^1_{k+1})) = -Ax(\tau^1_{k+1}), \lim_{x(\tau^1_{k+1}) \to 0} a^u(x(\tau^1_{k+1}), v(\tau^1_{k+1})) \to 0.$$

*Proof* (3c) applies the optimal transformation (Sec. 4) and relations (6.3),(6.3a). •

*Proposition* 6.3.

(1)-Holding constraint (6.4) for the considered conjugated Hamilton Eqs. brings the connections of complex conjugated eigenvalues $((\lambda_i(\tau), \lambda_j(\tau)))$ of this equation's operator $A^v(\tau)$ at each moment $\tau = \tau^1_k$ to the forms

$$\text{Re}\,\lambda_i(\tau^1_k) = \text{Re}\,\lambda_j(\tau^1_k),\ \text{Im}\,\lambda_i(\tau^1_k) = \text{Im}\,\lambda_j(\tau^1_k) = 0; \qquad (6.8)$$

(2)-Joint solutions of Eqs (6.4) and (6.4a) under applying control $v(\tau^o_k) = -2x(\tau^o_k)$, starting at $\tau^o_k$, *indicates the moment* $\tau^1_k$ when the solution of the constraint equation approaches the process' punched locality, and control $v(\tau^o_k)$ should be turned off, with transferring the extremal movement to a random process at the following moment $\tau_{k+1}$;

(3)-At the moment of turning the constraint off: $(\tau^1_k + o) \to \tau_{k+1}$, the control joins both real eigenvalues (6.9) of the conjugated $A^v(\tau)$ unifying these two eigenvalues:

$$\text{Re}\,\lambda_{ik}(\tau^1_k + o) = \text{Re}\,\lambda_i(\tau^1_k) + \text{Re}\,\lambda_k(\tau^1_k) \qquad (6.9)$$

and, hence, joins each $i, j$ dimensions, corresponding these conjugated eigenvalues and leading to *unification of the model's conjugated processes*;

(4)-Each moment $\tau^1_k$ of the control's turning off is found from the completion of the constraint Eq. (6.4) in the form (6.9) (at this moment) under the starting control $v(\tau^o_k) = -2x(\tau^o_k)$;

(5)- Within time interval $\tau^1_k - \tau^o_k = t_k$, the dynamic movement, approximating the diffusion process with *minimal entropy functional on the trajectories*, moves to the locality of imposing constraint with *a maximal information speed of decreasing this functional; and by reaching this locality it minimizes this speed.*

At the moment $\tau^1_k$, the extremal approaches the diffusion process with *a maximal probability*.

*Proof* (1).Specifying constraint equations (6.4), (6.4a) at the ending moment $\tau^1_k$ of interval $t_k, k = 1,...,m$, we have

$$X_i = A_i(x_k + v_k)(2b_i)^{-1}, X_k = A_k(x_i + v_i)(2b_k)^{-1}, r_{ik} = E[(x_i + v_i)(x_k + v_k)] = r_{ki}, \qquad (6.9a)$$

$$\frac{\partial X_i}{\partial x_k} = (2b_i)^{-1} A_i, A_i = -(r_i)^{-1} b_i;\ \frac{\partial X_k}{\partial x_i} = (2b_k)^{-1} A_k, A_k = -(r_k)^{-1} b_k.$$

Substitution these relations in the constraint's Eq.:

$$\frac{\partial X_i}{\partial x_k} = -2 X_i X_k = \frac{\partial X_k}{\partial x_i} \qquad (6.9b)$$

leads to $E[X_i X_k] = 1/4 r_{ii}^{-1} r_{ik} r_{kk}^{-1}, r_{ii} = E[x_i^2(t)], r_{kk} = E[x_k^2(t)]$ - at the central part of Eq. (6.9b), and to $E[\frac{\partial X_i}{\partial x_k}] = 1/2 r_{ii}^{-1}, E[\frac{\partial X_k}{\partial x_i}] = 1/2 r_{kk}^{-1}$ -on the left and right sides of the above Eq.

Suppose, imposing the constraint brings the connection of conjugate variables in the form $E[X_i X_k] \neq 0$, which holds true at the moment $\tau^1_k$ of the constraint starts.

It is seen that, at any finite auto-correlations $r_{ii} \neq 0, r_{kk} \neq 0$, this connection is holding only if $r_{ik}(\tau^1_k) \neq 0$.

The vice versa: $E[X_i X_k] = 0$ leads to $r_{ik} = 0$ at any other moments when the constrain is absent.

Joining the above relations for the central and both sides of (6.9a), we get

$$1/2 r_{ii}^{-1} = 1/2 r_{ii}^{-1} r_{ik} r_{kk}^{-1} = 1/2 r_{kk}^{-1}, or\ r_{ii}(\tau^1_k) = r_{kk}(\tau^1_k) = r_{ik}(\tau^1_k).$$

This means that the initial auto-correlations $r_{ii}(t), r_{kk}(t)$ become the mutual correlations $r_{ik}(t)$ at the moment $t = \tau_k^1$ of the constraint imposing, and random states $\tilde{x}_i(t), \tilde{x}_k(t)$ of the dimensions $i, k$ become connected.

Taking the derivations: $\dot{r}_{ii}(\tau_k^1) = \dot{r}_{kk}(\tau_k^1)$ leads to $b_i(\tau_k^1) = b_k(\tau_k^1)$, and then by multiplying this equation on $r_{ii}(\tau_k^1)^{-1} = r_{kk}(\tau_k^1)^{-1}$ we get $\dot{r}_{ii}(\tau_k^1) r_{ii}(\tau_k^1)^{-1} = \dot{r}_{kk}(\tau_k^1) r_{kk}(\tau_k^1)^{-1}$ which, by following (6.9a), leads to

$$A_i(\tau_k^1) = A_k(\tau_k^1), \qquad (6.10)$$

or for the matrix's eigenvalue $i, k$, considered at the same moment $\tau_k^1$, we get

$$\lambda_i(\tau_k^1) = \lambda_k(\tau_k^1). \qquad (6.10a)$$

Thus, imposing the constraint brings the *connection of model's processes* in the form (6.10a).

For a related complex conjugated eigenvalues, corresponding to a Hamilton Eqs. of a dynamic movement: $\lambda_i(\tau_k^1) = \alpha_i(\tau_k^1) + j\beta_i(\tau_k^1), \lambda_i^*(\tau_k^1) = \alpha_i(\tau_k^1) - j\beta_i(\tau_k^1)$,

satisfying (6.10a) in the form $\lambda_i(\tau_k^1) = \lambda_i^*(\tau_k^1)$, we come to

$$\operatorname{Im}\lambda_i(\tau_k^1) = \beta_i(\tau_k^1) = \operatorname{Im}[\lambda_i^*(\tau_k^1)] = -\beta_i(\tau_k^1), 2\beta_i(\tau_k^1) = 0, \operatorname{Im}\lambda_i(\tau_k^1) = 0, \qquad (6.10b)$$

and $\qquad \operatorname{Re}\lambda_i(\tau_k^1) = \operatorname{Re}\lambda_i^*(\tau_k^1). \qquad (6.10c)$

These prove (6.9) in (1).

*Proof* (2). After applying the control at the moment $\tau_k^o$ to both (5.5a) and (6.4) and using Eqs (6.6cb):

$$b(t)\frac{\partial X(t)}{\partial x(t)} + A(\tau_k^o, t)x(\tau_k^o, t)(2 - \exp(A(\tau_k^o)t))^T X^T(t) + 1/2 a^u(\tau_k, \tau_k^o, t)(2b(t)^{-1}(a^u(\tau_k, \tau_k^o, t)))^T = -\frac{\partial S}{\partial t}(t), (6.11)$$

the model moves to complete condition (5.10), decreasing the entropy functional with the maximal entropy speed (6.11), and when the constraint is turning on at the moment $\tau_k^1$, we get

$$\frac{\partial X(\tau_k^1)}{\partial x(\tau_k^1)} + 2A(\tau_k^o, \tau_k^1)x(\tau_k^o, \tau_k^1)(2 - \exp A(\tau_k^o)\tau_k^1)X^T(\tau_k^1) = 0, \qquad (6.11a)$$

the entropy speed reaches its minimum.

Thus, completion of (6.11a) *indicates* the moment $\tau_k^1$ when the solution of the constraint equation approaches the process' punched locality, and control $v(\tau_k^o)$ should be turned off, transferring the extremal movement to a random process at the following moment $\tau_{k+1}$. It's seen that satisfaction of any (6.10),(6.10a-c) actually follows from imposing the constraint (starting at $\tau_k^o$), which by the moment $\tau_k^1$ reaches the $(\tau_{k+1} - o)$ locality. Therefore, the conditions (6.10),(6.10a-c) are indicators of the (6.11a) completion, and hence, can be used to find the moment $\tau_k^1$ of turning the control off.

*Proof* (3). As it follows from Col.6.1, both real eigenvalues (6.8) should be transformed to a real component of diffusion matrix (according to (6.3)) after turning off the constraint. This requires completion of both (6.10) and (6.11a).

*Proof* (4)- (5). The impulse control action at the moment $\tau_k^1$ transfers the dynamic movement to the punched locality and activates the constraint in the form (5.23) on at this locality. At this moment, dynamic movement approaches the diffusion process with minimal entropy speed and the corresponding maximal probability. As it follows from Prop. 5.2, state $x(\tau_k^1)$ holds a *dynamic* analogy of random state $x(\tau_{k+1} - o)$ with a maximal probability, which is used to form a stepwise control (6.2), starting a next extremal segment; while both stepwise controls (forming of the impulse control) connects segments between the boundary points. The extremals provide a dynamic approximation of the diffusion process and allows its modeling under a currently applied optimal control. •

*Proposition 6.4*

(1)-Math expectation of the IPF derivation in (5.5a) (taken on each extremal by the moments $\tau = \tau_k$) and the matrix's $A(\tau)$ eigenvalues for each of the model dimension $i = 1,...,n,$ are connected via the equations

$$E[\frac{\partial \tilde{S}}{\partial t}(\tau)] = 1/4 Tr[A(\tau)],  \qquad (6.12)$$

$$A(\tau) = (\lambda_i(\tau)), \lambda_i(\tau_k) = 4E[\frac{\partial \tilde{S}_i}{\partial t}(\tau_k)], i = 1,...n, k = 1,...m. \qquad (6.12a)$$

The *proof* follows from substituting to (5.6) the relations (6.1), (6.3) that leads to Eq

$$E[-\frac{\partial \tilde{S}}{\partial t}(\tau)] = 1/2 E[(x(\tau) + v(\tau))^T A(\tau)^T (2b)^{-1} A(\tau)(x(\tau) + v(\tau))],$$

from which at $A(\tau) = -b(\tau) r_v^{-1}(\tau)$ we get (6.10); and for each $i$-dimensional model's eigenvalue with $\tau = \tau_k$ we come to (6.10a), where for a stable process $A(\tau) < 0$ •

(2)- The IPF for a total process, with $A(\tau) = -1/2 \sum_{i=1}^{n} \dot{r}_i(\tau) r_i^{-1}(\tau), (r_i) = r$, measured at the punched localities, acquires the form

$$I_{x_t}^p = -1/8 \int_s^T Tr[\dot{r} r^{-1}] dt = -1/8 Tr[\ln(r(T)/\ln r(s)], (s = \tau_o, \tau_1,..., \tau_n = T). \qquad (6.13)$$

(3)-An elementary entropy increment $S_i^\delta$ between the nearest segments' time interval $(\tau_k^i - o, \tau_k^i + o)$ corresponding to the cutoff (Sec.3) is

$$S_i^\delta = -1/8 \int_{\tau_k^i - o}^{\tau_k^i + o} r_i^{-1}(t) \dot{r}_i(t) dt = -1/8 \ln[r_i(\tau_k^i + o)/r_i(\tau_k^i - o)]. \qquad (6.13a)$$

Considering an approximation of these moments by those at the end of a previous segment: $\tau_k^i - o = \tau_k^{1i}$ and the beginning of a following segment: $\tau_k^i + o = \tau_k^{oi+1}$, we get

$$S_i^\delta = 1/8 [\ln r_i(\tau_k^{1i}) - \ln r_i(\tau_k^{oi+1})], \qquad (6.13b)$$

where the correlations are taken sequentially in the times course, from the moment $\tau_k^{1i}$ to the moment $\tau_k^{oi+1}$.
This quantity of information, evaluated in (Sec.3a): $S_i^\delta \cong 0.5 Nats$, delivers the process' hidden information. •

## 7. The model's information invariants.

*Proposition* 7.1. The constraint's information invariant $E[\Delta \tilde{S}_{ic}] = \Delta S_{ic} = inv$ (5.21a) during interval $t_k$ of applying controls $v(\tau_k^o) = -2x(\tau_k^o)$ leads to the following three invariants (for each model's dimension $i = 1,...,n$ and $k = i$):

$$\lambda_i(\tau_k^o) \tau_k^o = inv_1 = i_1 \qquad (7.1)$$
$$\lambda_i(\tau_k^1) \tau_k^1 = inv_2 = i_2 \qquad (7.1a)$$
$$\lambda_i(\tau_k^o) t_k = inv_3 = i_3, \qquad (7.1b)$$

and to their connections in the forms
$$i_1 = 2i_2, \qquad (7.2)$$
$$i_2 = i_3 \exp i_3 (2 - \exp i_3)^{-1}, \qquad (7.2a)$$

where $\lambda_i(\tau_k^o)$ and $\lambda_i(\tau_k^1)$ are complex eigenvalues of the model matrix $A = (\lambda_i)_{i=1}^n$, taken at the at the moments $\tau_k^o$ and $\tau_k^1$ of the segment's time interval $t_k$ accordingly. • Detailed proof is in (Lerner 2010a).

The *proof* uses the invariant information contribution by constraint (6.8c) in form:

$$\Delta S_{ic} = -1/2 \int_{\tau_k^o}^{\tau_k^1} \lambda_i(t) dt = -1/2 (\lambda_i(\tau_k^1)\tau_k^1 - \lambda_i(\tau_k^o)\tau_k^o) = inv \qquad (7.3)$$

which, according to Col 5.1 is connected to the Hamiltonian at this moment: $-2H_i(\tau_k^1)\tau_k^1 = 1/2 \lambda_i(\tau_k^1)\tau_k^1$ in form

$$\lambda_i(\tau_k^1)\tau_k^1 - \lambda_i(\tau_k^o)\tau_k^o = inv = -\lambda_i(\tau_k^1)\tau_k^1, \qquad (7.4)$$

which leads to $\lambda_i(\tau_k^1)\tau_k^1 = 1/2 \lambda_i(\tau_k^1)\tau_k^1 = inv$. \qquad (7.4a)

This proves both (7.1), (7.1a), and also (7.2). To prove (7.1b) we use the eigenvalues' function

$$\lambda_i(\tau_k^o)\exp(\lambda_i(\tau_k^o)t_k)[2 - \exp(\lambda_i(\tau_k^o)t_k)] \qquad (7.5)$$

following from (6.6b) after applying the control $v_i(\tau_k^o) = -2x_i(\tau_k^o)$ at $A = (\lambda_i), i = 1,...,k,...,n$.

Multiplying both sides of (7.5) on $t_k$ and substituting invariant (7.4) we obtain

$$\lambda_i(\tau_k^1)t_k = \lambda_i(\tau_k^o)t_k \exp(\lambda_i(\tau_k^o)t_k)[2 - \exp(\lambda_i(\tau_k^o)t_k)] \; . \qquad \bullet \qquad (7.6)$$

Considering the real eigenvalues for the above complex eigenvalues at each of the above moments:
$\operatorname{Re} \lambda_i(\tau_k^o) = \alpha_i(\tau_k^o)$, $\operatorname{Re} \lambda_i(\tau_k^1) = \alpha_i(\tau_k^1)$, we come to real forms of invariant relations in (7.1-7.2):

$$\alpha_i(\tau_k^1)\tau_k^1 = \mathbf{a}_i, \; \alpha_i(\tau_k^o)\tau_k^o = 2\mathbf{a}_i, \; \alpha_i(\tau_k^o)t_k = \mathbf{a}_{io}, \qquad (7.7)$$

Applying relations (7.1), (7.1a-b) for the complex eigenvalues' imaginary parts:
$\operatorname{Im} \lambda_i(\tau_k^1) = \beta_{ik}$, $\operatorname{Im} \lambda_i(\tau_k^o) = \beta_{iko}$ brings the related imaginary invariants:

$$\beta_i(\tau_k^1)\tau_k^1 = \mathbf{b}_i, \; \beta_i(\tau_k^o)t_k = \mathbf{b}_{io}. \qquad (7.7a)$$

## 8. The evaluations of the model's information contributions by the invariants

Following (7.7), we evaluate information, delivered at moment $\tau_k^o$, and information, produced by the dynamics at moment $\tau_k^o$, via the related invariants for each dimension $i = 1,...,n$, consequently:

$$\Delta S_i^{\delta o} = \alpha_i(\tau_k^o)\tau_k^o = 2\mathbf{a}_i, \Delta S_i^{do} = \alpha_i(\tau_k^1)\tau_k^1 = \mathbf{a}_i. \qquad (8.1)$$

Then, connection of these invariants (7.2a): $\mathbf{a}_i = -\mathbf{a}_{io} \exp(-\mathbf{a}_{io})(2 - \exp(-\mathbf{a}_{io}))^{-1}$ determines invariant evaluating information spent on information dynamics: $\Delta S_{io}^o = \mathbf{a}_{io}$, while the ratio

$$\Delta S_{io}^o / \Delta S_{io}^{do} \cong |\mathbf{a}_{io} / \mathbf{a}_i| = \exp(-\mathbf{a}_{io})(2 - \exp(-\mathbf{a}_{io}))^{-1}, \qquad (8.2)$$

at $\gamma \to (0.0 - 0.6)$ takes values between $3.28 - 2.63$, and at $\gamma \cong 0.5$ holds $\exp(-\mathbf{a}_{io})(2 - \exp(-\mathbf{a}_{io}))^{-1} \cong 3$, which satisfies the balance Eq. $\Delta S_{io}^o = |\mathbf{a}_{io}| \cong 3\mathbf{a}_i$. Thus, *the balance is* estimated at $\mathbf{a}_{io} \cong -0.7$, $\gamma \cong 0.5$.

Invariants $\mathbf{b}_{io}$ and $\mathbf{b}_i$ evaluate the model segment's information, generated by the eigenvalues imaginary components. Eq. (6.10b,c) impose the restriction on the solution of (6.1) during each interval $t_k$ of extremal segment, which we express through the equation for invariant $\mathbf{a}_{io}$ and the starting eigenvalue's ratio $\gamma_i = \beta_i(\tau_k^o)/\alpha_i(\tau_k^o)$:

$$2\sin(\gamma_i \mathbf{a}_{io}) + \gamma_i \cos(\gamma_i \mathbf{a}_{io}) - \gamma_i \exp(\mathbf{a}_{io}) = 0. \qquad (8.3)$$

This equation allows us to find *function* $\mathbf{a}_{io} = \mathbf{a}_{io}(\gamma_i)$ for each starting eigenvalue $\alpha_{iko} = \alpha_i(\tau_k^o)$, satisfying constraint (6.4), and also get $\gamma_i(\alpha_{iko})$, indicating dependence of $\gamma_i$ on identified $\alpha_{iko}$.

At each fixed $\gamma_i = \gamma_{i*}$, invariant $\mathbf{a}_{io}(\gamma_{i*})$ determines quantity of information needed to develop the information dynamic process approximating initial random process with a maximal probability.

Using relation (6.6d), Prop.6.2, we get the covariation (correlation) at the moment $\tau_k^1$:

$r_i(\tau_k^1) = r_i(\tau_k^o)[2 - \exp \mathbf{a}_{io}(\gamma_i)]^2, \mathbf{a}_{io}(\gamma_i) < 0,$  (8.4)

expressed via information invariant $\mathbf{a}_{io}(\gamma_i)$ and covariation $r_i(\tau_k^o)$ at the moment $\tau_k^o$. From (8.4) it follows that at $\mathbf{a}_{io}(\gamma_i = 0.5) \cong \ln 2$, the covariation $r_i(\tau_k^o)$ by the moment $\tau_k^1$ increases in $\cong 2.25$ times. Between these moments, the ratio of information (8.2) increases in $\cong 3$, or the expention of above covariation in $\cong 2.25$ times is an equivalent of the increase of information in $\cong 3$ times during the same time interval. This means that a hidden information, delivered through the correlation $r_i(\tau_k^o)$, by the moment $\tau_k^1$ will be compensated by that consumed through the information dynamics.

Actually, hidden *information, generated in the window* through the correlation (6.13b), is integrated by the moment $\tau_k^o - o$, and at the moment $\tau_k^o$ is transferred in the invariant form $\mathbf{a}_i(\gamma)$ by the segment's starting control to the segment's dynamics. The applied controls, delivering $2\mathbf{a}_i(\gamma)$ units of information, allow to increase this information in $\cong 3$ times, *through the optimal dynamics, which expands the initial hidden information up to the moment* $\tau_k^1$.

Otherwise, the moment $\tau_k^1$ is determined by the condition of reaching the above equilibrium.

Between these moments, the relative speed of extension of this correlations determines the information speed of dynamic process: $1/2 \dot{r}_{it} r_{it}^{-1} = \alpha_{it}$ (8.4a), according to (7.3). The information contribution of this extension, evaluated by

$$\int_{\tau_k^o}^{\tau_k^1} \alpha_{it} dt = 1/2 \int_{\tau_k^o}^{\tau_k^1} \dot{r}_{it} r_{it}^{-1} dt = \ln(r_i(\tau_k^1)/r_i(\tau_k^o)),$$  (8.4b)

(and therefore by (6.13)) equals to the macrodynamic information contribution up to the moment $\tau_k^1$ of imposing the constraint. Moreover, above information speeds of the correlation's expansion (8.4a), by reaching the moment $\tau_k^1$, provide satisfaction of the constraint Eq. (6.9a,b), particularly, in the form

$$b_i \frac{\partial X_i}{\partial x_i}(\tau_k^{1i}) = -2(X_i)^2 = 1/4 \dot{r}_{it} r_{it}(\tau_k^{1i}).$$  (8.4c)

<u>Comments 8.1.</u>

Let us find the *connection* between the real and imaginary eigenvalues for a regular matrix $A(\tau_k^1) = A(\tau_k^o) \exp(A(\tau_k^o) t_k)$, which would be necessary to convert its complex eigvalues into a real eigenvalue during time interval $t_k = \tau_k^1 - \tau_k^o$, starting at moment $\tau_k^o$. By requesting $\lambda_i(\tau_k^1) = \lambda_i^*(\tau_k^1)$, (8.5), we have

$\lambda_i(\tau_k^1) = [\alpha_i(\tau_k^o) + j\beta_i(\tau_k^o)] \exp[(\alpha_i(\tau_k^o) + j\beta_i(\tau_k^o)) t_k] = [\alpha_i(\tau_k^o) - j\beta_i(\tau_k^o)] \exp[(\alpha_i(\tau_k^o) - j\beta_i(\tau_k^o)) t_k] = \lambda_i^*(\tau_k^1).$

And then we get $[\alpha_i(\tau_k^o) + j\beta_i(\tau_k^o)]/[\alpha_i(\tau_k^o) - j\beta_i(\tau_k^o)] \exp(2 j\beta_i(\tau_k^o) t_k) = \lambda_i^*(\tau_k^1)$, from which it follows

$[\alpha_i(\tau_k^o) - j\beta_i(\tau_k^o)]^2 / [\alpha_i(\tau_k^o)^2 + \beta_i(\tau_k^o)^2] = [\cos(2\beta_i(\tau_k^o) t_k) - \sin(2\beta_i(\tau_k^o) t_k)],$

$[\alpha_i(\tau_k^o)^2 - \beta_i(\tau_k^o)^2] / [\alpha_i(\tau_k^o)^2 + \beta_i(\tau_k^o)^2] = \cos(2\beta_i(\tau_k^o) t_k)$

$2\alpha_i(\tau_k^o) \beta_i(\tau_k^o) / [\alpha_i(\tau_k^o)^2 + \beta_i(\tau_k^o)^2] = -\sin(2\beta_i(\tau_k^o) t_k).$

Then, using indications $\beta_i(\tau_k^o)/\alpha_i(\tau_k^o) = \gamma_i$, $\beta_i(\tau_k^o) t_k = b_{io}$, we come to Eqs

$(1 - \gamma_i^2)/(1 + \gamma_i^2) = \cos 2b_{io}, -2\gamma_i/(1 + \gamma_i^2) = \sin 2b_{io},$

from which we get

$\text{Arc} \cos(1 - \gamma_i^2)/(1 + \gamma_i^2) = 2b_{io} = \text{Arc} \sin[-2\gamma_i/(1 + \gamma_i^2)].$  (8.5a)

The main solutions of (8.5) in diapason $-\pi/2 \leq \text{Arc} \sin 2b_{io} \leq \pi/2, 0 \leq \text{Arc} \cos 2b_{io} \leq \pi$ satisfy to the Eq. $(1 - \gamma_i^2)/(1 + \gamma_i^2) = -2\gamma_i/(1 + \gamma_i^2)$ which has a solution $\gamma_i = 1 \pm \sqrt{2}$. At $\alpha_i(\tau_k^o) < 0, \beta_i(\tau_k^o) < 0$, and the admissible $|\gamma_i| \in (0.0 - 1.0)$, it brings $\gamma_{io} \cong -0.4142$ and $2b_{io}(\gamma_{io}) = \pi/4, b_{io}(\gamma_{io}) = \pi/8$.  (8.6)

This means that constraint (6.11) in the form (8.5) imposes the connection $|\beta_i(\tau_k^o)| = 0.4142|\alpha_i(\tau_k^o)|$ on the initial imaginary and real components the eigenvalues and therefore on the solutions of the conjugated Eqs.

Let us check if this solution satisfies to the Eq. (8.5) for the matrix's imaginary component:

$$Im[\alpha_i(\tau_k^o) + j\beta_i(\tau_k^o)]\exp[(\alpha_i(\tau_k^o)t_k + j\beta_i(\tau_k^o)t_k]$$
$$= j\exp(\alpha_i(\tau_k^o)t_k)[\alpha_i(\tau_k^o)\cos(\beta_i(\tau_k^o)t_k) + \beta_i(\tau_k^o)\sin(\beta_i(\tau_k^o)t_k)] = 0.$$

We get $\cos(b_i) + \gamma_i \sin(b_i) = 0, \gamma_i = -tgb_i$, which at $b_i = \pi/8$ brings the initial $\gamma_i = -0.4142 = \gamma_{io}$.

The real component:

$$Re[\alpha_i(\tau_k^o) + j\beta_i(\tau_k^o)]\exp[(\alpha_i(\tau_k^o)t_k + j\beta_i(\tau_k^o)t_k]$$
$$= \exp(\alpha_i(\tau_k^o)t_k)[\alpha_i(\tau_k^o)\cos(\beta_i(\tau_k^o)t_k) + \beta_i(\tau_k^o)\sin(\beta_i(\tau_k^o)t_k)]$$

leads to equality

$$\alpha_i(\tau_k^o)\exp(b_i/\gamma_i)[\cos(b_i) + \gamma_i \sin(b_i)] = \alpha_i(\tau_k^o)[\cos\pi/8 - 0.4142\sin\pi/8]\exp(-0.948),$$

which allows finding both $\alpha_i(\tau_k^1) \cong 0.386\alpha_i(\tau_k^o)$ and time interval $t_k = 1/8\pi/0.4142|\alpha_i(\tau_k^o)|$ for any initial real $\alpha_i(\tau_k^o)$ identified at beginning of this interval.

At $\mathbf{a}_{io}^1 = \alpha_i(\tau_k^o)t_k = b_{io}/\gamma_{io}$, we get $\gamma_{io} = 0.4142$, while from Eq (8.3), obtained for the controllable model, we have $\mathbf{a}_{io}(\gamma_{io} = 0.4142) \cong 0.73$, clearly seeing their difference at the same restricted $\gamma_{io} = 0.4142$. •

Let us find optimal $\gamma_{io}$ corresponding to an *exteme* of function of the information invariant:

$$\mathbf{a}_i(\gamma_i) = \mathbf{a}_{io}(\gamma_i)\exp\mathbf{a}_{io}(\gamma_i)(2 - \exp\mathbf{a}_{io}(\gamma_i))^{-1}, \tag{8.7}$$

which would be necessary to develop an *optimal* dynamic process approximating the measured information process.

Using a simple indications $\mathbf{a}_{io}(\gamma_i) = y(x), \mathbf{a}_i(\gamma_i) = z(y(x))$, we get

$$etrz = dz/dy(dx) = [(2 - \exp y)][(dy/dx)\exp y + (dy/dx)y\exp y[(2 - \exp y)]^{-2}$$
$$- y\exp y(-(dy/dx)\exp y)[(2 - \exp y)]^{-2} = 0;$$
$$(dy/dx)(1 + y)\exp y + (dy/dx)y\exp 2y[(2 - \exp y)]^{-1} = 0, \exp y \neq 2,$$
$$(1 + y) + y\exp y[(2 - \exp y)]^{-1} = 0, (dy/dx) \neq 0, y \neq 0.$$

We come to Eq. $2(1 + y) = \exp y$, whose solution $y \cong -0.77 = \mathbf{a}_{io}(\gamma_{io})$ (8.8)

allows us to find, from (8.3), $\gamma_i$ equivalent to (8.8a): $\gamma_{io} \to 0$, and using (8.6) to get $\mathbf{a}_i(\gamma_{io}) \cong 0.23$. (8.8a)

This is a minimal $\mathbf{a}_i(\gamma_{io})$ in diapason of admissible $|\gamma_i| \in (0.0 - 1.0)$, while at

$$\gamma_{io} \to 0, \beta_i(\tau_k^o) \to 0, \text{ at } \mathbf{b}_{io} \to 0, \mathbf{a}_{io} \to \max. \tag{8.8b}$$

Minimal invariant $\mathbf{a}_i(\gamma_{io})$ delivers maximal information from a cutoff random process during interval of applied impulse control and provides a maximum information $\mathbf{a}_{io}(\gamma_{io})$ for optimal dynamic process, corresponding to implementation of the minimax principle.

The information measure of the above invariants $\mathbf{a}_{io}(\gamma_{io})$ and $\mathbf{a}_i(\gamma_{io})$ correspond to 1.1 bit and 0.34 bits accordingly.

Information needed by a stepwise control, starting this *i*-segment, is evaluated by $\mathbf{a}_{ic}(\gamma_{io})$; another stepwise control, turning the constraint off, requires the same information $\mathbf{a}_{ic}(\gamma_{io})$.

The invariant information, spent on the dynamics $\mathbf{a}_{io}(\gamma_{io}) = \alpha_i(\tau_k^o)t_k^i$, covers information needed by both stepwise control and that produced by the time interval end : $\Delta S_{io}^{do}$, leading to balance Eq

$$\Delta S_{io}^{do} = \mathbf{a}_i(\gamma_{io}) = \mathbf{a}_{io}(\gamma_{io}) - 2\mathbf{a}_{ic}(\gamma_{io}). \tag{8.9}$$

This information should be extracted from a random process, which is evaluated by

$$I_{so}^i = \mathbf{a}_i(\gamma_{io}). \tag{8.9a}$$

For the considered optimal process this information is estimated by $\mathbf{a}_i(\gamma_{io}) \cong 0.23$, and a total $\mathbf{a}_i(\gamma_{io}) \cong 0.77$.

Then information, needed for a single stepwise control in this optimal process:

$$\mathbf{a}_{ic}(\gamma_{io}) = 1/2(\mathbf{a}_{io}(\gamma_{io}) - \mathbf{a}_i(\gamma_{io})), \tag{8.9b}$$

is estimated by $\mathbf{a}_{ic}(\gamma_{io}) \cong 0.27 Nat$.

We assume that impulse control's information, consisting of the two stepwise controls:

$$2\mathbf{a}_{ic}(\gamma_{io}) \cong 0.54 Nat, \tag{8.9c}$$

provides an extractror (observer), which also generates the information dynamics, evaluated by the invariants (8.9a).

Comments 8.2.

An observered process could also deliver the impulse information.

It has shown (Lerner 1973, 2008) that imposition of two and more physical processes generates an impulse control action.

For example, thermo-electric, thermo-kinetic, electro-chemical and other cross-phenomena are associated with their mutually controllability each other, which is analogous to the impulse jumpwise interactions.

These cross-phenomena (with related impulses) are sources of physical information in a natural objective observer (Sec. 2.12) and Lerner 2012. An interaction of a subjective observer with its external information can also bring an impulse control. A natural objective observer, at a quantum level, possesses quantum impulse control (Zurek 2007).

Any of the above impulse controls could bring information during an interval of observation, evaluated by analogy with invariant relations $2\mathbf{a}_{ic}(\gamma_i)$. •

At unknown control information, it can be approximated by a dynamic invariant in the form

$$2\mathbf{a}_{ic}(\gamma_i) \cong \mathbf{a}_{io}^2(\gamma_i). \tag{8.10}$$

For the optimal dynamics we get $\mathbf{a}_{io}^2(\gamma_{io}) \cong 0.593 Nat$ and $2\mathbf{a}_{ic}(\gamma_{io})$ close to (8.9c). In average, we have

$$\mathbf{a}_{ic}(\gamma_i) \approx \mathbf{a}_i(\gamma_i) \approx 1/3\mathbf{a}_{io}(\gamma_i). \tag{8.10a}$$

Information, extracted during a time interval $\delta\tau_k^i$, is measured by the correlations (6.13) at some initial moments $\tau_k^i = \tau_o^i, \tau_k^i + \delta\tau_k^i = \tau_o^i + \delta\tau_o^i$, in the form

$$I_s = -1/8 Tr[\ln(r_{ij}(\tau_o^i, \tau_o^i + \delta\tau_o^i) / r_{ij}(\tau_o^i))], i,j = 1,....,n, \ I_s = \sum_{i=1}^{n} I_s^i. \tag{8.11}$$

Following (Levy 1965), we estimate ratio

$$\tau_o^i / [(\tau_o^i + \delta\tau_o^i) - \tau_o^i)] = r_i^{1/2}(\tau_o^i) / r_i^{1/2}(\tau_o^i, \tau_o^i + \delta\tau_o^i) \text{ by}$$

$$\delta\tau_o^i / \tau_o^i = [r_i(\tau_o^i, \tau_o^i + \delta\tau_o^i) / r_i(\tau_o^i)]^{1/2}. \tag{8.12}$$

The estimation holds true specifically for random state $\tilde{x}_i(\tau_o^i), \tilde{x}_i(\tau_o^i + \delta\tau_o^i)$ between the some moments $(\tau_o^i, \tau_o^i + \delta\tau_o^i)$ of the process's cutoff, and for the correlation $r_i(\tau_o^i, \tau_o^i + \delta\tau_o^i)$ (that supposed to be cut off between these moments).

Information $I_s^i$, obtained during time interval $\delta\tau_o^i$, is evaluated by the invariant (following from (8.1)):

$$I_s^i = \alpha_i(\tau_o^i)\delta\tau_o^i = 2\mathbf{a}_i. \tag{8.13}$$

Joining (8.11) and (8.12), we get

$$[r_{ij}(\tau_o^i + \delta\tau_o^i) - r_{ij}(\tau_o^i)] / r_{ij}(\tau_o^i) = \exp[-16|\mathbf{a}_i(\gamma_i)|], \tag{8.14}$$

and then using (8.13), we have

$$\delta\tau_o^i / \tau_o^i = [\exp(-8|\mathbf{a}_i(\gamma_i)|], \tag{8.14a}$$

which determines a relative interval of applying impulse control.

For example, to get a minimal $\mathbf{a}_i(\gamma_{io}) \cong 0.23$ (8.8a) we need interval $\delta\tau_o^i / \tau_o^i \cong 0.1588 \sec$.

A stepwise control, starting the extremal segment at moment $\tau_1^{io} = \delta\tau_o^i + o$, converts eigenvalue $\alpha_i(\tau_o^i)$ (in 8.13) to
$\alpha_i(\tau_1^{io}) = |\alpha_i(\tau_o^i)|$.

Then (8.14) holds form
$I_s^i = 2|\mathbf{a}_{io}(\gamma_i)|\delta\tau_o^{*i} = 2|\mathbf{a}_i(\gamma_i)|, \mathbf{a}_{io}(\gamma_i) = \alpha_i(\tau_o^i)t_1^i, \delta\tau_o^{*i} = \delta\tau_o^i / t_1^i$,

and we get interval $\delta\tau_o^i$, relative to $t_1^i$, being estimated by the invariants:

$$\delta\tau_o^{*i} = |\mathbf{a}_i(\gamma_i)| / |\mathbf{a}_{io}(\gamma_i)|. \tag{8.15}$$

This means that getting constant quantity information $\mathbf{a}_{io}(\gamma_{io})$, needed for optimal dynamics, requires applying *variable* time intervals $\delta\tau_k^{*i}$ of extracting such information $I_s^i$, which would depend on the correlations, which are cutting off from the random process.

The moment $\tau_o^i$ of applying stepwise control, we evaluate by a stepwise function

$$\tau_o^i(t) = \begin{cases} 0, t < \tau_o^i \\ (1, t = \tau_o^i) \end{cases} \text{ from which it follows } \delta\tau_o^i / \tau_o^i = \delta_{oi}^*(\tau_o^i = 1) = \delta\tau_o^i. \tag{8.16}$$

For a minimal $I_{so}^i = 2|\mathbf{a}_i(\gamma_{io})|$, optimal time interval $\delta\tau_o^{*i} = \mathbf{a}_i(\gamma_{io}) / \mathbf{a}_{io}(\gamma_{io}) \cong 0.3$, which at $\delta\tau_o^i \cong 0.1588 \sec$ brings evaluation of the related segment's time interval's: $t_1^i \cong 0.16 / 0.3 \cong 0.533 \sec$.

At known invariants, we get the ratio of the eigenvalues at the end and beginning of any segment's time interval:

$$\alpha_i(\tau_k^1) / \alpha_i(\tau_k^o) \cong \mathbf{a}_i(\gamma_{io}) / \mathbf{a}_{io}(\gamma_{io}), \mathbf{a}_i(\gamma_{io}) / \mathbf{a}_{io}(\gamma_{io}) \cong 0.3, \tag{8.17}$$

which allows us to estimate both $\alpha_i(\tau_k^1) \cong 0.3\alpha_i(\tau_k^o)$ and time interval $t_k = |\mathbf{a}_{io}(\gamma_{io})| / |\alpha_i(\tau_k^o)|$ for any initial real $\alpha_i(\tau_k^o)$ identified at beginning of this interval. For the considered optimal $t_1^i$, we may predict

$$\alpha_i(\tau_o^i) = |\mathbf{a}_{io}(\gamma_i)| / t_1^i \cong 1.313. \tag{8.18}$$

For any $\gamma_i \neq \gamma_{io}$, invariant $\mathbf{a}_i(\gamma_i)$ is found from known external information $I_s^i$:
$I_s^i = 2\mathbf{a}_i(\gamma_i)$.

Using (8.7), we find the related invariant $\mathbf{a}_{io}(\gamma_i)$ and then $\gamma_i$ based on (8.3).

After that, following (8.9b) is found the needed control information $\mathbf{a}_{ic}(\gamma_i)$.

Thus, all information invariants follow from the measured external information.

## 9. The procedure of the dynamic modeling and a prediction of the diffusion process

We assume that the observed random process is modeled by Markov diffusion process with the considered stochastic equation, and the EF-IPF relations hold true.

With a nonobservable stochastic Eq, its diffusion component can be identify through correlations of the observed random process: $r = E[\tilde{x}_t \tilde{x}_t^T]$ with $\dot{r} = 2b = \sigma\sigma^T$, while its drift function is defined through the identification Eq (6.3), Sec.6.

The considered *model* is initiated by a starting stepwise control, applied at the moment $\tau^o = \tau_o + \delta\tau_o$; $x(\tau_o)$ are the object's initial correlations

$$r_o = E[x(\tau_o)x^T(\tau_o)], \text{ and/or } b_o = 1/2\dot{r}_o, \tag{9.1}$$

which determine nonrandom initial $x(\tau_o) = \{x_i(\tau_o^i)\}$ at

$$x_i(\tau_o^i) = \pm c_i |r_i(\tau_o^i)|^{1/2} \tag{9.2}$$

for the starting control
$$v_i(\tau_i^o) = -2x_i(\tau_o^i),  \quad (9.3)$$
or the control in the form (6.2a)):
$$u(\tau_o, \tau^o) = b(\tau_o) r(\tau_o)^{-1} v(\tau^o). \quad (9.4)$$
Relation (9.2) follows from the model's Eqs (5.8) in the form
$$\dot{x} = 2bX, X = 1/2r^{-1}x, \quad (9.5)$$
which for each dimension holds $dx_i / x_i dt = 1/2 dr_i / r_i dt$, and after integrating leads to (9.2).

*The procedure starts* with capturing from the diffussion process its correlation
$$r = E[\tilde{x}_t(t)\tilde{x}_t^T(t+\delta t)], \; r = (r_{ij}), i, j = 1,..n \quad (9.6)$$
during fixed time intervals $t = \tau, \delta t = \delta \tau, \tau = (\tau_k^i), k = 0,1,2,...$ , beginning with the process' initial moments $\tau_o = (\tau_o^i)$.
The correlation provides the considered cut off of diffusion process at these moments and the process' segments, which are approximated by the segments of related macrodynamic process.

According to (8.11, 8.12, 8.14a) these correlations measure a quantity of information obtained on each $\delta\tau_o^i$:
$$I_s = -1/8Tr[\ln(r_{ij}(\tau_o^i, \tau_o^i + \delta\tau_o^i)/r_{ij}(\tau_o^i))] , \; I_s = \sum_{i=1}^n I_s^i \quad (9.7)$$
which for any $i$-the dimension is evaluated by information invariant (8.13).

Applying (8.13, 8.14) consists of computing $r_i (\tau_o^i, \tau_o^i + \delta\tau_o^i)$ from (9.6) at fixed $r_i(\tau_o^i)$ and $\tau_o^i$ with sequential increasing increments of $\tau_o^i + \delta t$ until $\delta t = \delta\tau_o^i$ will satisfy (8.14).

This connects the evaluation of relative interval $\delta\tau_o^i / \tau_o^i = \delta_{oi}^*$ by information $\mathbf{a}_i$ in the form (8.14a).

Using (9.6-9.7) and (8.14a), we get jointly $\delta\tau_o^i$, $\mathbf{a}_i$ and then $|\alpha_i(\tau_o^i)|$ from (8.13).

At the moment $(\tau_o^i + \delta\tau_o^i) = \tau_i^o$, step-up control (9.3) (being a right side of impulse control) transfers the eigenvalue $\alpha_i(\tau_o^i)$ (computed according to (9.7)) to a dynamic model. This indicates extraction of information (9.7) from a random process during interval $\delta_{oi}^*$ and start the segment's macrodynamic process, which, at the end of time interval $t_1^i$ is turned off by a step-down control $v_i(\tau_i^1) = -2x_i(\tau_i^1)$ (as a left-side of the impulse control for the next segment).

Both stepwise controls provides the considered cutoff, but, in reality, during a *finite* time interval $\delta_{oi}^*$.

The segment's initial nonrandom states $x(\tau_o) = \{x_i(\tau_o^i)\}$ can be find using the correlation function in (9.2).
It allows to apply the initial optimal control (9.3) to all primary extremal segments, starting optimal dynamics simultaneously at the moment $\tau^o = (\tau_i^o)$ on these segments.

Since correlations $r_i(\tau_o^i, \tau_o^i + \delta\tau_o^i)$ for different process' dimensions would be diverse, as well as $I_s^i$, the invariants $\mathbf{a}_i$, intervals $\delta\tau_o^i$, and eigenvalues $\alpha_i^k(\tau_o^k)$ will also be dissimilar.

That's why the different primary time intervals $t_i^1(\tau_o^i), i = 1,...,n$ of the information dynamics on each segment will form a sequence, being automatically ordered in the time course:
$$t_1^1(\tau_o^1), t_2^1(\tau_o^2), t_3^1(\tau_o^3),...t_i^1(\tau_o^i),...,t_n^1(\tau_o^n). \quad (9.8)$$
This allows us to arrange the related $\tau_o = (\tau_o^i)$ and $\alpha_i(\tau_o^i)$ in the ordered series.

For example, the ordered arrangement would bring a hierarchical decrease (or increase) of these eigenvalues: $|\alpha_m(\tau_o^m)|, m = 1,..,n$, where each successive *m*-th eigenvalue of such series is priory unknown.

Moments $t_i^1(\tau_o^i)$ will determine the moments $\tau_1 = (\tau_1^i)$ of measuring correlation (9.6) and through repeating the procedure next sequence of $t_i^2(\tau_1^i)$ will be obtained:

$$t_1^2(\tau_1^1), t_2^2(\tau_1^2), t_3^2(\tau_1^3),...,t_i^2(\tau_1^i),...,t_n^2(\tau_1^n). \tag{9.9}$$

And so on, each previous $t_i^{k-1}(\tau_{k-1}^i)$ will determine the following $t_1^k(\tau_k^1), t_2^k(\tau_k^2), t_3^k(\tau_k^3),...,t_i^k(\tau_k^i),...,t_n^k(\tau_k^n)$.

The ranged series of eigenvalues at the end of each $\alpha_i(\tau_i^k)$ is needed for sequential joining them in a hierarchical information network (IN), modeling information structure of whole *n*-dimensional dynamic system (Sec. 2.1).
Therefore, the summarized procedure consists of :

1. Simultaneous start whole system by the applying impulse controls at moments $(\tau_o, \tau^o), \tau^o = \tau_o + \delta\tau_o$ with computation of the process' correlations during their time intervals for each dimension;
2. Finding the quantities of information, selected at these time intervals, the information invariant, and the related eigenvalues of the potential information dynamics on each extremal segment;
3. Finding the initial dynamic states (at the end of interval $\tau^o$) and starting the stepwise controls at these moments, which, with the known eigenvalues, initiate dynamic process at beginning of each segment;
4. Finding the time intervals of these dynamics on each segment during the random process' time course and arranging the segment's ending eigenvalues in the terms of their decreasing;
5. Turning off the stepwise control (at the segment's end) and starting measuring the process' correlations on each moment $\tau_1 = (\tau_1^i)$ determined by the end of each $t_i^1(\tau_o^i)$;
6. Finding new information invariants, the related eigenvalues, initial dynamic states, and turning *on* a new stepwise control on the next segments;

7. Turning *off* the previous stepwise controls (after finding $t_i^2(\tau_1^i)$) and measuring the process' correlations, finding the information invariants, eigenvalues, initial dynamic states, and then starting the following controls, which activate information dynamics on each segment.

The procedure allows us to approximate each segment of the diffusion process by each segment of information dynamics with a maximal probability on each segment's trajectory, thereafter transforming the random process to the probability's of equivalent dynamic process during a real time course of both processes.

Periodic measuring the random process' correlations allows us to renovate each segment's dynamics by the current information, extracted from the random process during its time course, at the predictable moments when dynamics coincides with the stochastics with maximal probability, closed to one.

In such approximation, each random segment's quantity information is measured by equivalent quantity of dynamic information, expressed by invariant $\mathbf{a}_{oi}$. This invariant depends on the real eigenvalues, measured by the correlations at the moments, when these quantities are equalized, and, therefore, both segments' hold the information equivalence above.

The estimation of each subsequent correlations by correlations at the end of each previous segment $r_i(t \to \tau_k^{1i})$ (which follows from preservation of diffusion component of the random process at the considered cut off), can *predict* each next segment's dynamics through the eigenvalues $\alpha_i(\tau_{k+1}^i) \approx 1/2 r_i^{-1}(\tau_k^{1i}) \dot{r}_i(\tau_k^{1i})$ computed at the segment's ending moment $\tau_k^{1i}$, which estimates the eigenvalues at the moment $\tau_{k+1}^i$.

With the optimal invariant (Sec.8), quantity of dynamic information for each segment will be constant while a width of the control impulse $\delta\tau_k^i$ would depends on the quantity of external information.

Since interval between these impulses $t_k^i$, measured by invariant $\mathbf{a}_{io}(\gamma_{io})$, will be also constant, such an encoding of a random process (by $\mathbf{a}_{io}(\gamma_{io})$, $t_k^i$, and $\delta\tau_k^i$), is described by a varied pulse-width modulation (Fig.2a).

At any fixed $\mathbf{a}_{io} \neq \mathbf{a}_{io}(\gamma_{io})$, both interval $t_k^i$ and pulse-width $\delta\tau_k^i$ will be different, and encoding is described by a pulse-amplitude modulation with a varied time ( Fig.2b). At a fixed external information, various $\delta\tau_k^i$ will generate different $t_k^i$, which is described by pulse-time modulation (Fig.2c).

Therefore, the information, collected during a window, *explicitly determines dynamic process* along each extremal segment through: the extremal's initial conditions (found from (9.2)) and the control starting on them (9.3); the initial eigenvalue (found from (8.13), where the invariant and the window's duration are found jointly from (8.12-8.14a)); and the duration of the dynamic process (9.8),(9.9).

All considered above operations sequentially repeat themselves during the time course of the diffusion process, which is modeled by the extremal's dynamics with maximal probability.

The impulse control also joins the extremal segments in their *sequential chain* with building a cooperative process.

*The specific* of the considered optimal process *consists of the computation of each following time interval* (where the identification of the object's operator will take place and the next optimal control is applied) *during the optimal movement under the current optimal control, formed by a simple function of dynamic states.*

In this *optimal dual* strategy, the IPF optimum predicts each extremal's segments movement not only in terms of a total functional path goal, but also by setting at each following segment the renovated values of this functional's controllable drift and diffusion (identified during the optimal movement), which currently correct this goal.

*The automatic control system, applying dynamic feedback with time delay $t_k$, which is dependable on an object's current information, has been designed, patented, and implemented in practice* (Lerner1961, 1989).

## 10. The law of information dynamics:

$$\dot{x}_i = 2b_i X_i, X_i = \frac{\partial \Delta S_i}{\partial x_i}, \qquad (10.1)$$

following from the Hamilton equation (5.2,5.3), defines a dynamic speed $\dot{x}_i = I_i$ as an information flow $I_i$ for the macroprocess' information states $x(t) = \{x_i(t)\}, i = 1,...,n$, which is a carrier of information (novelty) within each segment. Each local increment of an *information potential* $\Delta S_i$ is limited by the segment's total information potential $\Delta S_{ic}(x_i(t_k))$ at the points $x_i(t = \tau_k^1)$ of stopping the information movement on the segment's end.

According to (10.1), a local *flow* of information $I_i$, carried by a dynamic state's information speed, is *proportional* to a gradient of information potential $X_i$ and a dispersion $b_i$ of each random state's ensemble $\tilde{x}_i(t)$.

This relation models their *causal-consecutive relationship*. At a fixed $X_i$ and an increased dispersion, this novelty is distributed with growing speed, which corresponds to a dissemination of novelties, when, at growing $b_i$, the sets of their carriers (the states' ensembles with the process' probabilities) are spreading.

The information law (10.1) is extended on the space of moving collective and cooperating systems with the explicitly defined geometrical borders (Lerner 208, 2010a, 2011).

We may also specify the information force $X_\delta = \frac{\partial \Delta S_\delta}{\partial x}$, using $\partial x = 1/2 c_o r^{-1/2} \partial r(\tau)$ for state $x(\tau) \cong c_o r^{1/2}(\tau)$ and $\frac{\partial \Delta S_\delta}{\partial r} = -1/8 r^{-1}$ (from (6.11a)). Then we obtain

$$X_\delta = \frac{\partial \Delta S_\delta}{1/2 c r^{-1/2} \partial r} = -1/4 c_o r^{-1/2}. \quad (10.2)$$

This is a finite information force, which keeps the states of process bounded (Sec.1), counterbalancing to the destructive force that arises at dissolving correlations, at $r \to 0$ (Sec.3a,(3.7b)). The constraint in the form

$$\frac{\partial X_i}{\partial x_k} = -2 X_i X_k = \frac{\partial X_k}{\partial x_i}, \quad (10.3)$$

at the moment of its imposing, binds both information forces and their gradients of this macrodynamics process.

Information, generated by a random process on each window, is converted to a flow of the information macrodynamic process, which is measured and renovated at each predictable moment of starting and ending the window. A current flow, interrupted at the windows, is assembled in an information network (IN) (Part 2), which organizes the assembled flow's portion in a hierarchical chain, determined by both quantity and quality of the carried information.

Therefore, the VP allows establishing not only information connections between the changes (Sec.1), identified by specific sources and related information measures, but also *finding the regularities* of these connections in the form of a complex *causal-consecutive relationship*, carried by EF-IPF analytics through both the flow's chain and the IN.

## 11. Specifics of the information process at a window. About quantum dynamics at a locality of the window

The VP solution provides a *piece-wise approximation* of the random process' trajectories by the extremal trajectories of the optimal process with a maximal probability. It requires dividing the random process on its pieces, which are modeled by the corresponding segments of extremal trajectories. Selection of both the pieces of random trajectories and related segments of extremal trajectories provide the stepwise controls (SP1 and SP2, Fig.1b), applied simultaneously to the random process and its dynamic model during the same time intervals. This leads to minimizing both EF and IPF.

While the IPF is minimized along the segments' dynamics, reaching its minimax by the segments' end, the EF gets its minimax under the applied impulse control IC(SP1, SP2) (Sec.3a,Prop.5.2).

Along the segment, the IPF is minimized with a maximal speed of decreasing this functional (Col.5.2), and by the end of dynamic movement at the moment $\tau_k^{i1} - o_t(\tau_k^{i1})$), when a solution of dynamic constraint: $x_i(t_k^i)$ approaches the dynamic boundary points of diffusion process $x_i(\tau_k^{i1} - o)$ (Sec.5), this speed reaches its minimum.

A jump of this speed takes place at $o(\tau_{k+1}^i)$-locality of diffusion process between states $\tilde{x}_i(\tau_{k+1}^i - o), \tilde{x}_i(\tau_{k+1}^i), \tilde{x}_i(\tau_{k+1}^i + o)$, measured by the increment $S_{\tau_{k+1}}^{\delta u_i} = E[\delta \varphi_{si}^{\mp}]_{\tau_{k+1}}$, $\tau_k = (\tau_k^i), i = 1,...n$, where both EF minimax' and IPF minimax' information measures coincide.

The jump, indicating the coincidence, brings a delta probability distribution during interval $o(\tau_{k+1}^i)$ with a maximal functional probability of the transformation (1.2).

Stepwise control SP1 $v_i(\tau_k^{i1})$, terminating constraint at the moment $\tau_k^{i1} + o_1(\tau_k^{i1})$, indicates an "edge" $o_1(\tau_k^{i1})$ of the dynamics when this deterministic control *intervenes* in a random boundary. During the switch $o_1(\tau_k^{i1})$ of control $v_i(\tau_k^{i1})$ the dynamic motion is *transferred on* the diffusion process' boundary at the moment $\tau_k^{i1} + o_1(\tau_k^{i1})$.

At reaching a proximity, the dynamic state $x_i(\tau_k^{i1} - o)$ *meets* the related Markovian state $\tilde{x}_i(\tau_{k+1}^i - o)$, attracted to this boundary at the moment $\tau_k^{i1} + o_1(\tau_k^{i1}) = (\tau_{k+1}^i - o)$, with a possibility of their interaction.

In stochastic differential Eq.(1.2), this control, acting on the process drift, cuts it, transforming Markov diffusion process $\tilde{x}_i(t < (\tau_{k+1}^i - o))$ to Brownian process $\tilde{x}_i(t \geq (\tau_{k+1}^i - o))$ (with zero drift).

Stepwise control SP2 $v_i(\tau_{k+1}^{io})$, which intervenes in this random boundary at the moment $\tau_{k+1}^i$, implements the probability transformation $P[\tilde{x}_i(\tau_{k+1}^i)] \to P[\tilde{x}_i(\tau_{k+1}^i + o)]$ (by analogy with control (3.1) transforming (3.7a), Sec3), and captures the related dynamic border state during time interval $o(\tau_{k+1}^{io})$ of the control's switch, when it starts the next segment dynamics. The process' states, cut off by this control, have *maximal* entropy at this cutting moment.

The time window $\delta_o(\tau_{k+1}^i)$ between the nearest extremal's segments:

$$\delta_o(\tau_{k+1}^i) = o_1(\tau_k^{i1}) + \delta(\tau_{k+1}^i) + o(\tau_{k+1}^{io}), \qquad (11.1)$$

is formed by the two intervals of switching both stepwise controls and the interval $\delta(\tau_{k+1}^i)$ of collecting information from Markovian stochastics. Such $\delta(\tau_{k+1}^i)$-window, selected by impulse control $\delta v_i[\delta(\tau_{k+1}^i)]$ at the delta probability distribution, is the most informative source of maximal information, chosen among all available on the window:

$$I_i^s = \delta S_i[x_i \delta(\tau_{k+1}^i)], \qquad (11.2)$$

which is needed for generation of each following segment's dynamics; while each collected information subsequently renovates itself in recurring actions.

That integral information measures the selected sequence of random microstates of initial diffusion process (Sec.3), being enclosed in a process correlation, holds the same information measure in its invariant form $\mathbf{a}_i$ (9.3), where invariant $\mathbf{a}_i$ is a macroscopic information measure of the hidden microstates' sequence, which the correlation binds.

The selected portion of diffusion process, as a priori process, being transformed to the following extremal segment of a posteriori process, approximates this portion with a maximal probability and measures it by $\mathbf{a}_i$. Total information, generated at $\delta_o(\tau_{k+1}^i)$ is measured by invariant $\mathbf{a}_{io} \cong 3\mathbf{a}_i$ that summarizes the measured information (11.2) with that needed for the both controls, which implement this transformation to macrodynamics and then back to stochastics.

In the Markov random filed, random ensemble of states, chosen from the delta distribution and possessing a *maximum* of the information *distinction simultaneously*, should be *opposite* directional. That's why when the step-up optimal control

$$v_i(\tau_{k+1}^{io}) = -2x_i(\tau_{k+1}^i) \qquad (11.3)$$

intervenes in the opposite directional random processes $\tilde{x}_{t+}(\tau_{k+1}^i), \tilde{x}_{t-}(\tau_{k+1}^i)$ (11.3a), control function (11.3) activates doubling each of these states with the opposite signs and selecting the opposite directional copies

$x_+(\tau_{k+1}^i), x_-(\tau_{k+1}^i)$ (11.3b) of the random process $\tilde{x}_{t+}(\tau_{k+1}^i), \tilde{x}_{t-}(\tau_{k+1}^i)$ to form the opposite step-up controls

$$v_{o1} = -2x_+(\tau_{k+1}^i), v_{o2} = -2x_-(\tau_{k+1}^i). \qquad (11.4)$$

By the following moment $\tau_{k+1}^{io}$, the opposite directional step-up actions initiate the *dynamic* processes $x_t(x_+(t), x_-(t))$, composed of the parallel conjugated information dynamic processes:

$$x_{1k+1,2k+1}^{io}(t) = \mp 2x(\tau_{k+1}^{io})[\cos(\alpha_{k+1}^{io}t) \pm j\sin(\alpha_{k+1}^{io}t)], \qquad (11.5)$$

whose time interval depends on both initial information speed $\alpha_{k+1}^{io}$ and starting amplitude $|2x(\tau_{k+1}^{io})|$ of these dynamics.

The initial conditions $-(\pm 2x(\tau_{2o}))$ are selected by controls (11.4) at the above intervention, while their information speed is identified through the measured *relative* dispersion function of Markov process:

$$\alpha_{k+1,\tau}^{i} = b_{k+1,\tau}^{i} (\int_{k+1,\tau} b_{k+1,t}^{i} dt)^{-1}, b_{k+1,t}^{i} = 1/2\dot{r}_{k+1,t}^{i}, \qquad (11.6a)$$

where $r_{k+1,t}^{i}$ is correlation of the information process, delivered on the widow's locality $\delta(\tau_{k+1}^{i})$.

The control transforms information speed (11.6a) to the process' dynamics as its eigenvalue

$$\alpha_{k+1}^{io} = -\alpha_{k+1,\tau}^{i}. \qquad (11.6b)$$

Markov diffusion (11.6a), drives this information speed and limits it by maximal acceptable diffusion noise, which restricts maximal frequency of the available information spectrum.

These process' dynamics will be ad joint at the moment $\tau_{k+1}^{1i}$ by the end of time interval $t_{k+1}^{i}$, which depends on the obtained cutoff portion of information (11.2).

Specifically, information $\hat{S}[\tilde{x}_{t}(\tilde{x}_{t+}(\tau_{k+1}^{i}), \tilde{x}_{t-}(\tau_{k+1}^{i}))]$, extracted under control (11.2), generates complex information function $\tilde{S}(x_{1k+1,2k+1}^{io}(t)) = [\tilde{S}(x_{1k+1}^{io}(t)), \tilde{S}(x_{2k+1}^{io}(t))]$ of the conjugated processes $x_{1k+1}^{io}(t)$ and $x_{2k+1}^{io}(t)$:

$$\tilde{S}(x_{1k+1}^{io}(t)) = \tilde{S}_{a}(t) + j\tilde{S}_{b}(t), \tilde{S}(x_{2k+1}^{io}(t)) = \tilde{S}_{a}(t) - j\tilde{S}_{b}(t), \qquad (11.7)$$

which at the moment $\tau = \tau_{k+1}^{1i}$ of imposing the dynamic constraint, satisfy equality:

$$\tilde{S}_{a}(\tau_{k+1}^{1i}) = -\tilde{S}_{b}(\tau_{k+1}^{1i}). \qquad (11.7a)$$

This leads to converting total initial information in double real information $2\tilde{S}_{a}(\tau_{k+1}^{1i})$:

$$\tilde{S}(x_{1k+1}^{io}(\tau_{k+1}^{1i})) + \tilde{S}(x_{2k+1}^{io}(\tau_{k+1}^{1i})) = -\tilde{S}_{a}(\tau_{k+1}^{1i})(j-1) + \tilde{S}_{a}(\tau_{k+1}^{1i})(j+1) = 2\tilde{S}_{a}(\tau_{k+1}^{1i}). \qquad (11.7b)$$

Because of that, at $\tilde{S}_{a}(\tau_{k+1}^{1i}) = \mathbf{a}_{i}(\tau_{k+1}^{1i})$, total real dynamic information $\mathbf{a}_{id}(\tau_{k+1}^{1i})$, generated at this moment by control (11.4), being applied during $t_{k+1}^{i}$, is

$$2\tilde{S}_{a}(\tau_{k+1}^{1i}) = 2|\mathbf{a}_{i}(\tau_{k+1}^{1i})| = |\mathbf{a}_{id}(\tau_{k+1}^{1i})|, \mathbf{a}_{id}(\tau_{k+1}^{1i}) = 2\mathbf{a}_{i}(\tau_{k+1}^{1i}). \qquad (11.8)$$

A dynamic movement within an extremal segment, satisfying the model's Hamilton Eqs (Secs.5,6), holds a pair complex conjugated eigenfunctions $\lambda_{i,j}(t) = \alpha_{i}(t) \pm j\beta_{i}(t)$. By the moment $\tau_{k+1}^{i1}$ of ending the segment's dynamic movement, such a pair "collapses" (satisfying (11.8) at the same moments of time) with joining the pair into a single real eigenfunction:

$$\lambda_{i}(\tau_{k+1}^{i1}) + \lambda_{j}(\tau_{k+1}^{i1}) = 2\alpha_{k+1}^{i}(\tau_{k+1}^{i1}), \qquad (11.8a)$$

which is related to a jump of the real information speed.

The delivered information, needed for the information dynamics, consists of the following parts:(1)- information, collected from the observations $\mathbf{a}_{i}(\tau_{k+1}^{i})$, which is transferred to $\mathbf{a}_{i}(\tau_{k+1}^{io})$; (2)-the starting control's information $\mathbf{a}_{i1c}(\tau_{k+1}^{io})$; (3)-the stepwise control information $\mathbf{a}_{i2c}(\tau_{k+1}^{i1})$, generating the turning-off action, when (11.7,11.7a are satisfied. Summing all external information contributions leads to balance Eq

$$\mathbf{a}_{i}(\tau_{k+1}^{io}) + \mathbf{a}_{i1c}(\tau_{k+1}^{io}) + \mathbf{a}_{i2c}(\tau_{k+1}^{i1}) \cong \mathbf{a}_{io}. \qquad (11.9)$$

Starting information $\mathbf{a}_i(\tau_{k+1}^{io}) + \mathbf{a}_{i1c}(\tau_{k+1}^{io})$ compensates for information, needed for the conjugated dynamics $\mathbf{a}_{id}(\tau_{k+1}^{1i})$, thereafter satisfying the balance Eq of the Hamiltonian information dynamics:

$$\mathbf{a}_i(\tau_{k+1}^{io}) + \mathbf{a}_{i1c}(\tau_{k+1}^{io}) = \mathbf{a}_{id}(\tau_{k+1}^{1i}). \tag{11.9a}$$

Under the control action, requiring information $\mathbf{a}_{i2c}(\tau_{k+1}^{1i})$, total dynamic information $\mathbf{a}_{id}(\tau_{k+1}^{1i})$ is entangled in the ad joint conjugated processes (11.7a) at the moment $\tau_{k+1,o}^i = \tau_{k+1}^{1i} + o$, following $\tau_{k+1}^{1i}$. From (11.9, 11.9a) it follows Eq.

$$\mathbf{a}_{id}(\tau_{k+1,o}^i) + \mathbf{a}_{i2c}(\tau_{k+1}^{1i}) \cong \mathbf{a}_{io}, \tag{11.9b}$$

which, while preserving invariant $\mathbf{a}_{io}$, includes the entangled information $\mathbf{a}_{id}(\tau_{k+1,o}^i) = \mathbf{a}_{ien}(\tau_{k+1,o}^i)$, evaluated by $\mathbf{a}_{ien}(\tau_{k+1,o}^i) \cong 2\mathbf{a}_{i1c}(\tau_{k+1}^{oi})$, and free information $\mathbf{a}_{i2c}(\tau_{k+1}^{1i})$ that enables both local and non-local attracting cooperative actions. Free information, carrying the control's cut-down action, is able to interact with random portion of the external information by cutting down its correlation, while a following step-up action provides cut up the correlation, attraction and extraction of its information. The attracting action, currying maximal information, enables selecting from the random observation a most likely sample of the random ensemble, having opposite directional states (11.3a), the related dynamic states (11.3b) and then renovates the information dynamics through repeating the above procedure.

Each impulse's cutoff information, measured by discrete portion $\mathbf{a}_i(\tau_{k+1}^i)$ of information functional (11.2), is a *code*, selected from observed process according to minimax, while the set-up control copies this information and transforms it to the equivalent dynamic information during $t_{2k}$, which depends on $\mathbf{a}_i(\tau_{k+1}^i)$.

Let us analyze the components of time interval (11.1) at transferring the process from $k$ to $k+1$ segment.

The entanglement at $\tau_k^{i1} - o_t(\tau_k^{i1})$ approaches boundary dynamic state $x_i(\tau_k^{i1} - o)$, located at the end of each extremal segment, prior to the stepwise control (SP1) $v_i(\tau_k^{i1})$ termination, which is predicted by VP. Such entanglement precedes interaction of the already entangled (dynamic) state with a random state at the boundary, which potentially changes this state. Hence, interval $o_t(\tau_k^{i1})$, preceding the constraint termination, should be a finite but infinitively small (in classical dynamics), because the constraint's termination leads to disappearance of the boundary dynamic states $x_i(\tau_k^{i1} - o)$, which should emerge within the time $\tau_k^{i1} - o$ of holding the constraint. This leads to $\tau_k^{i1} - o \cong \tau_k^{i1} - o_t(\tau_k^{i1}), o \cong o_t(\tau_k^{i1})$.

From other aspect, the entangled state $x(\tau_k^{i1})$ should be transferred to the dynamic border's state $x_i(\tau_k^{i1} - o)$ during same time interval $o_1(\tau_k^{i1})$, following the turning constraint off. In the transformations

$$x(\tau_k^{i1} - o) \to x(\tau_k^{i1}) \to x(\tau_k^{i1} + o(\tau_{k+1}^i)), \tag{11.10}$$

it brings $o(\tau_{k+1}^i) \cong -o$, which is impossible from a classical causal-consequence relationship (11.10).

That is why should be held the following relations

$$x(\tau_k^{i1} - o) \cong x(\tau_k^{i1}) \cong x(\tau_k^{i1} + o(\tau_{k+1}^i)) \text{ at } o \cong o_t(\tau_k^{i1}) \cong o(\tau_{k+1}^i) \to 0, \tag{11.11}$$

which bring the entanglement infinitively closed to the random interaction on an edge of the window.

Moreover, because the moment of boundary point should coincide with the moment of holding the constraint, both of them are supposed to emerge *simultaneously.*

In opposite case, a classical *causal-consecutive relationship* in this locality (dynamic boundary first) would be lost.

Specific of the transformation is associated with a *localization* of classical information dynamic movement along an extremal in *quantum* information dynamics in the above proximity and with connection of these quantum *information* phenomena to stochastics at the *transformation of dynamic quantum locality to a locality of random movement*.

In (Wilde *et al* 2012) it is a *shared entanglement*, considered as a distribution or consumption of the information, measured in bits and qubits accordingly.

Function of action $\hat{S}[x_t]$ of the information dynamic process, defined through additive functional (2.4) and probability density (2.3) satisfies to an information form of Schrödinger Eq. (Lerner 2012b) with related probability measure for the conjugated wave functions and a Hamiltonian $\hat{H}$ of the conjugated eigenfunctions.

The constraint joins the eigenfunctions in the forms (11.8a) (6.10a-c), which corresponds to an entanglement of the information wave functions in a quantum locality (11.7a,b).

During the dynamic movement, preceding to imposing the constraint, the conjugated probability's amplitudes are defined on the solutions of the Hamilton Eqs (Sec.5), corresponding the conjugated extremals (11.5).

The quantum phenomena at $o_t(\tau_k^{i1})$ locality (before the entanglement of eigenfunctions at the moment of imposing the constraint), bring an *uncertain situation* when both the end of the extremal movement and its continuation are equally possible *simultaneously*. Since the constraint's imposition and termination involve both classical and quantum dynamics, such uncertainty connects them (by analogy with Schrödinger's cat (Shrödinger1935)) and affects the macrolevel's information dynamics. Because the entanglement is infinitively closed to the random interaction on the window, this uncertainty leads to a *simultaneous probability of both* observing the random window and not observing it, or *both* extraction of information at the window and *not* extraction it.

The constraint works as a bridge connecting classical information dynamics, stochastics, quantum dynamics, and macrodynamics.

Since the information form of Schrödinger Eq. is generated only during a quantum interval $o_t(\tau_k^{i1})$, the information wave functions, evaluated by the *quantum*-conjugated probabilities, appear during $o_t(\tau_k^{i1})$ and then are cooperating at the entanglement. Within the interval of dynamic movement $t_k = (\tau_k^{i1} - o) - \tau_k^{io}$ (starting the dynamic constraint at $\tau_k^{io}$ and imposing it at $\tau_k^{i1}$) each of the conjugated dynamic processes $x_i(t), x_j(t)$ (11.5) depends on its complex amplitude probabilities $P_i[x_i(\tau_k^{io})/x_i(\tau_k^{i1})], P_j[x_j(\tau_k^{io})/x_j(\tau_k^{i1})]$, as the macrodynamic analogies of the related quantum conjugated probabilities. After imposing the constraint at $\tau_k^{i1}$, the macro conjugated probabilities turn into the quantum probabilities, which are joining at the moment of entanglement, generating the bound hidden variables

$$x_i(\tau_k^{i1}) \Leftrightarrow x_j(\tau_k^{i1}). \qquad (11.12)$$

At a following moment $\tau_k^{i1} + o_1(\tau_k^{i1}) = (\tau_{k+1}^i - o)$, the impulse control's step-down action cuts off the random process' correlation. The impulse $\delta$-control action assumes $\delta(\tau_{k+1}^i) \to 0$, which leads to $r_i[\delta(\tau_{k+1}^i)] \to 0$, hence, classical information consumption within this interval (11.2) will disappear. That is why, each impulse control consists of two stepwise actions with a *finite* time intervals $o_1(\tau_k^{i1})$ and $o(\tau_{k+1}^{io})$ between them.

From that, it follows: the same way as dynamic model predicts appearance of the Markovian boundary with the states $x_i(\tau_k^{i1}) \cong \tilde{x}_i(\tau_{k+1}^i)$ and the moment $\tau_{k+1}^i \cong \tau_k^{i1}$, it also predicts a finite correlation $r_i[\delta(\tau_{k+1}^i)]$ at the end of each preceding extremal segment.

Quantum entanglement, at the end of a segment, connects a pair of its eigenfunctions, which leads to decreasing a dimension of the dynamic operator at this segment.

In a multi-dimensional process, after connecting the nearest segments, the subsequent segment's dynamic movement (by its end) is able to produce a new entanglement with further decreasing of components of the initial model's matrix.

Such sequential connections of the segments tend to cooperation of all eigenfunctions into a single one by the end of optimal movement, which leads to the process' *collectivization*.

However, each transfer from segment's dynamics to the window's stochastics at $a_i^u(u_i[\delta(\tau_{k+1}^i)]) \to 0$ is accompanied by a local instability and arising of a chaotic process (related to a white noise at $\tilde{x}_t^a = \int_{\delta(\tau_{k+1})} \sigma(v, \zeta_v) d\zeta_v$), which is associated with a maximum of minimal entropy of the optimal random process being transformed according to (4.6) and Sec.5 (Comments 5.1).

The applied control, which assembles potentially cooperating segments, is able to stabilize these random fluctuations.

Possessing the considered information invariants allows a prediction of each subsequent movement of the process based on information, which is collected at each previous window.

The impulse control's jump-wise quantum selection provides the following *specifics* of the considered *Quantum Information Dynamics* (QID).

The step-down control (that terminates the segment's dynamics by turning the dynamic constraint off) is applied to the diffusion process at the moment $\tau_{k+1,o}^i$, which cuts the process under this control action, initiating an *information path* from $\tilde{x}_i(\tau_{k+1,o}^i)$ to $\tilde{x}_i(\tau_{k+1,d}^i)$ through the control's jump-wise selection.

The following step-up control action, applied to the diffusion process at the moment $\tau_{k+1,d}^i$, ends to the process' controlled path and starts the dynamic process at the next extremal segment at a boundary point $x_i(\tau_{k+1}^i)$.

A selected *copy* of random states $-2\tilde{x}_i(\tau_{k+1,d}^i)$ is transferred to classical dynamic states $-2x_i(\tau_{k+1}^{io})$ by a deterministic step-up control action $v_i(\tau_{k+1}^{io}) = -2x_i(\tau_{k+1}^{io})$, which starts the next segment's dynamics.

Since the two opposite directional dynamic Hamiltonian movements (Sec.5) start at the same random locality, each of their step-up control is made of two states: $\tilde{x}_i(\tau_{k+1,d}^i)$ and its copy $-\tilde{x}_i(\tau_{k+1,d}^i)$ and then is formed through their differences:

$$-\tilde{x}_i(\tau_{k+1,d}^i) - \tilde{x}_i(\tau_{k+1,d}^i) = -2\tilde{x}_i(\tau_{k+1,d}^i), \text{ or } \tilde{x}_i(\tau_{k+1,d}^i) - (-\tilde{x}_i(\tau_{k+1,d}^i)) = 2\tilde{x}_i(\tau_{k+1,d}^i). \quad (11.13)$$

Each of these differences, being transferred to the equivalent classic states, serves as the related control: $-2x_i(\tau_{k+1}^{io})$ that starts a dynamic process with initial condition $x_i(\tau_{k+1}^{io})$, and the control $2x_i(\tau_{k+1}^{io})$ that starts the dynamic process with initial conditions $-x_i(\tau_{k+1}^{io})$.

Forming the optimal controls from a pair of identical but opposite states, leads to creation of two identical dynamic processes with opposite directions having the phase shift between them $\pi$ or $-\pi$ (which become entangled at $\tau_k^{i1}$).

In the IPF Minimax, only a limited random path between states [$\tilde{x}_i(\tau_{k+1,o}^i)$, $\tilde{x}_i(\tau_{k+1,d}^i)$], predicted by this principle's VP, are selected from an observed process (at its punched localities-windows), to satisfy a *maximum* entropy.

Information, collected through this local path, gives a start to a *minimal entropy path*, chosen by the dynamic model (an observer) to find the path is *ending states* (producing a minimum entropy) as a stable and the most *preferable* states.

Such *optimal chosen path* reflects the *current observations with maximum probability* and instructs the observer's further evolution to be encoded in the observer's structural genetics (Parts 2, 3).

The IPF minimax predicts the localization of *measurement,* when the observer's initiated measurement interacts with the environment, which entangles a measured Markovian path (at its quantum level) (see also Meyer *et all*, 2001), creating the mixed states of the quantum path, holding a corresponding quantum (QM) *probability* $\hat{P} = |\hat{\psi}|^2$, ($\hat{\psi}$ is a complex amplitude of a wave function), and a quantum (QM) *information measure* $\hat{H} = -Tr\hat{P}\ln\hat{P}$.

The information from this quantum IPF path determines the QID *quantum information*, which is selected at each punched locality of a random process. At the predicted moment $(\tau - o)$ of the entanglement, $\hat{P}[\hat{x}(\tau - o)]$ reaches a maximum, compared to the probabilities of related pure states $\hat{P}[\hat{x}(t)], t < (\tau - o)$.

According to the VP, the dynamic trajectory approximates the EF measured random process with a maximal probability (on the process random trajectories).

Hence, the selected information states satisfy the predicted maximum probability (in probability theory) but might not satisfy a quantum probability of chosen quantum states.

W.H.Zurek 2007 demonstrates that a predicted wave's information value encloses specific properties of a measured process with a *following jump-cut action* on the process.

The VP predicted control jump, acting at the moment $\tau_{k+1}^i$, discontinues (decohers) the measured process (at these punched locations) and selects $\tilde{x}_i(\tau_{k+1,o}^i)$ with the related the quantum state $\hat{x}_i(\tau_{k+1,o}^i)$ holding minimal probability for both $P[\tilde{x}_i(\tau_{k+1,o}^i)]$ and $P[\hat{x}_i(\tau_{k+1,o}^i)]$, while breaking symmetry of a mixed quantum wave at $\hat{x}_i(\tau_{k+1,o}^i - o)$.

Since, the control jump dissolves the correlations of the nearest states, these states become independent (orthogonal), and thus they could be under other (following) measurement. Moreover, randomness of a measured process assumes the existence of many random paths, which can be measured at the same moment.

However, de-correlation of the random process between the path states' locations $\tilde{x}_i(\tau_{k+1,o}^i)$ and $\tilde{x}_i(\tau_{k+1,d}^i)$ reduces the quantity of information bound by these states (Sec.3) (as a hidden information) and, thus, diminishes the future information outcomes from this process.

The measurement along the path, brought by the impulse control action, satisfies the definition of information (Sec.1) for each of the path's distinguished event.

The orthogonality of a current measurement of $\hat{x}_i(\tau_{k+1,o}^i)$ also leads to selection of the orthogonal (eigen) states, while such orthogonality diagonalizes the operator that provides this measurement, allowing the selection of the operator's real eigenvalues. This also leads to diagonal zing of the identified correlation matrix.

The step-up control's action on the state $\tilde{x}_i(\tau_{k+1,d}^i)$ replicates this state, reapplying it to the dynamic trajectory at the moment $\tau_{k+1}^{io}$. The step-down control's action at the moment $\tau_{k+1}^i$ following $(\tau_{k+1,o}^i - o)$ breaks symmetry of the dynamics at these points. The selected states hold a maximal probability of the random process, and their dual copies are transferred by the step-up controls to *classical dynamic macrostates* on the IPF extremal trajectories.

Both breaking quantum symmetry, following the entanglement, *and* the de-correlation of nearest random states of the initial correlated *random* process depict an *interactive effect* of random environment on the measured process.

The controls, implementing above operations, bind the measured process with its environment during the interactive dynamics and spends the quantity of information $\mathbf{a}_{io}^2(\gamma)$ on the process measurement, with transferring this information to the starting dynamic movements. These operations, working insight of an interactive process, link up randomness, quantum information with classical dynamics.

The quantum information dynamics (QID) at each $\delta(\tau_{k+1}^i)$ locality differ from physical quantum mechanics (QM), even though the related Hamilton Eqs in both dynamics, satisfying to corresponding variation conditions, are applied.

The first difference consists in the QID origin from quantum localities of stochastic process, whose entropy functional (EF) is transformed to the dynamic information path functional (IPF) using the information variation principle, which imposes the constraint on this transformation. At this VP transformation, an information process, measured by the EF, is transferring to the IPF dynamics process with aid of the impulse control, applied to both random process and its dynamic model, which implements the constraint imposition and selects the EF pieces.

Both QM and QID are able to predict the moments of the process measurement, whose information encloses the specific properties of a measured process with a *following jump-cut action* on the process for its next measurement.

Secondly, the collapse of wave functions in QIM, following from imposing the constraint on Markovian stochastics, leads to a sequential connection of the model's entangled movements (with joining eigenfunctions) in a collective motion (Part 2).

<u>Comments 11</u>. (1). Applying stepwise control SP1 on the drift in (2.1), will turn to zero the density measure in (3.7a): $p_s^{t-} \to 0$. This brings a minimal probability and *maximum information* at the punched locality, corresponding (Sec.2). The starting SP2 (at $p_s^{t+} \to 1$) transfers this boundary state (Sec.5, Prop.2) to a next extremal, implementing transformation (2.2) with entropy functional's measure (2.6).

(2). Above procedure proceeds under the assumption of imposing constraint (DC), which connects drift and diffusion of the controllable diffusion process. Actually, such imposition takes place only at moments $(\tau_k^{i1} - o_t(\tau_k^{i1}))$ when condition of imposing DC (5.9b, 6.9b) completes, and before SP1 control is turning DC off at $\tau_k^{i1}$.

Therefore, even though each starting SP2 initiates DC in both the controllable diffusion process and its dynamic model, the DC gets in action only at $(\tau_k^{i1} - o_t(\tau_k^{i1}))$ locality, where EF reaches its minimum. At the following time interval, where the drift is zero and diffusion is the same as that in the previous locality, this minimum is maximized and transferred to a next starting extremal segment. Hence, each dynamic segment starts with maximum of minimal information, which its Hamiltonian dynamics will preserve.

However, for unobserved stochastic Eq, its drift function is unknown, but an observer of the process can measure an incoming information flow (11.2) and select its maximum from a minimal information by the jump-wise action (Sec.3a)(at the moment of selection), which implements VP and therefore imposes the DC.

After that, for the constrained drift and diffusion, the SP2 control applies to $a^u$ (according to Eq.(6.1)), and matrix $A$ is identified by the correlations at the punched locality, following Secs.6, 8,9.

Hence, this jump-wise action produces also information invariant $\mathbf{a}_i$, which allows finding $\mathbf{a}_{io}$ and then implementing all procedure above. The selected information of the observed process' bound states is transferred to its dynamic information process with a maximal probability, therefore, implements the VP. In this procedure, the observed process' hidden *information* generates the information macroprocess, whose dynamics reveals the process' regularities. •

**Fig.1a.** Applying controls: IC (SP1, SP2) on the window

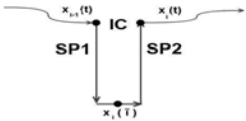

**Fig.1b.** Selection of the process' information portions (segments) and *assembling* them in a triple, *during* the initial flow's time-space movement

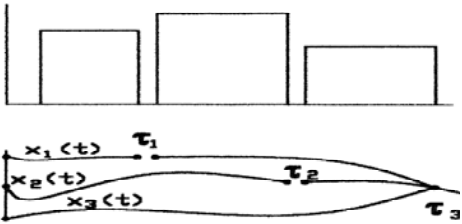

**Fig2.** Encoding random process applying the EF-IPF integral measures

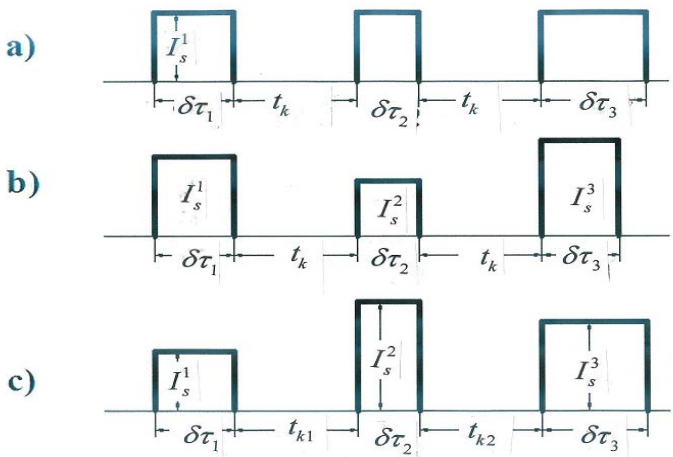